\documentclass[12pt]{article}
\pdfoutput=1
\usepackage{hyperref}
\usepackage{graphicx}
\usepackage{graphics}
\usepackage{dsfont}
\usepackage{epsfig}
\usepackage{amsmath,amssymb,amsthm,amscd}
\newcommand{\Tr}{\textrm{Tr}}

\def\ket{\rangle}
\def\bra{\langle}

\newcommand{\be}{\begin{equation}}
\newcommand{\ee}{\end{equation}}
\newcommand{\ba}{\begin{aligned}}
\newcommand{\ea}{\end{aligned}}

\numberwithin{equation}{section}

\begin{document}
\begin{titlepage}
{}~ \hfill\vbox{ \hbox{} }\break

\rightline{USTC-ICTS/PCFT-21-04}

\vskip 3 cm

\centerline{\Large 
\bf  
Free BMN Correlators  With More Stringy Modes  }

\vskip 0.5 cm

\renewcommand{\thefootnote}{\fnsymbol{footnote}}
\vskip 30pt \centerline{ {\large \rm 
Bao-ning Du\footnote{baoningd@mail.ustc.edu.cn}, 
Min-xin Huang\footnote{minxin@ustc.edu.cn}  
} } \vskip .5cm  \vskip 20pt 

\begin{center}
{Interdisciplinary Center for Theoretical Study,  \\ \vskip 0.1cm  University of Science and Technology of China,  Hefei, Anhui 230026, China} 
 \\ \vskip 0.3 cm
{Peng Huanwu Center for Fundamental Theory,  \\ \vskip 0.1cm  Hefei, Anhui 230026, China} 
\end{center}

\setcounter{footnote}{0}
\renewcommand{\thefootnote}{\arabic{footnote}}
\vskip 40pt
\begin{abstract}

In the type IIB maximally supersymmetric pp-wave background, stringy excited modes are described by BMN (Berenstein-Madalcena-Nastase) operators in the dual 
$\mathcal{N}=4$ super-Yang-Mills theory. In this paper, we continue the studies of higher genus free BMN correlators with more stringy modes, mostly focusing on the case of genus one and four stringy modes in different transverse directions. Surprisingly, we find that the non-negativity of torus two-point functions, which is a consequence of a previously proposed probability interpretation and has been verified in the cases with two and three stringy modes, is no longer true for the case of four or more stringy modes. Nevertheless, the factorization formula, which is also a proposed holographic dictionary relating the torus two-point function to a string diagram calculation, is still valid. We also check the correspondence of planar three-point functions with Green-Schwarz string vertex with many string modes. We discuss some issues in the case of multiple stringy modes in the same transverse direction. Our calculations provide some new perspectives on pp-wave holography.

\end{abstract}

\end{titlepage}
\vfill \eject

%%%%%%%%%%%%%%%%%%%%%%%%%%%%%%%%%%%%%%%%%%%%%%%%%%%%%%%%%%%%%

\newpage

\baselineskip=16pt

\tableofcontents

\section{Introduction}

The AdS/CFT correspondence \cite{Maldacena:1997re, Gubser:1998bc, Witten:1998qj} is a deep idea which relates two seemingly totally different theories, namely string theory or supergravity on AdS background and the $\mathcal{N}=4$ $SU(N)$ super-Yang-Mills theory. Although the correspondence has found flourishing applications in many topics, the precise quantitative tests of the holographic dictionary are mostly restricted to supersymmetry protected quantities in the supergravity approximation, such as the spectrum and correlation functions of BPS operators. Without an alternative effective method to handle string theory in the deeply stringy regime,  a common perspective is to simply take the super-Yang-Mills theory as a non-perturbative definition of AdS string theory at any finite coupling and energy scale, assumed to be valid unless otherwise convincingly explicitly contradicted.  

A particularly interesting avenue for progress in the precise tests of the holographic correspondence in the stringy regime is to take a Penrose limit  \cite{Penrose1976} of the type IIB $AdS_5\times S^5$ background. The geometry becomes a pp-wave background \cite{Blau:2001ne} with also maximal supersymmetry
\begin{eqnarray} \label{ppwave}
ds^2 = -4 dx^{+} dx^{-} -\mu^2 (\vec{r}^{~2} + \vec{y}^{~2} ) (dx^{+})^2 + d \vec{r}^{~2} + d  \vec{y}^{~2}, 
\end{eqnarray}
where $x^{+}, x^{-} $ are light cone coordinates, $\vec{r}, \vec{y}$ are 4-vectors, and the parameter $\mu$ is proportional to spacetime curvature as well as the Ramond-Ramond flux $F_{+1234}=F_{+5678}\sim \mu$. The free string spectrum can be solved in the light cone gauge using Green-Schwarz formalism similar to the flat space \cite{Metsaev:2001bj}. Berenstein, Maldacena and Nastase (BMN) proposed the holographic dual operators in the gauge theory for the stringy states, a type of near-BPS operators known as the BMN operators, and it was shown that the free string spectrum is reproduced by the planar conformal dimensions of these BMN operators \cite{Berenstein:2002jq}. On the field theory side, one takes a large R-charge limit, previously considered in the context of giant gravitons, or D-branes in the AdS space in \cite{McGreevy:2000cw, Hashimoto:2000zp, Balasubramanian:2001nh, Corley:2001zk}, and also in many subsequent literature e.g. \cite{Balasubramanian:2002sa, Balasubramanian:2004nb, Pasukonis:2010rv, Koch:2011hb}. The calculations on the field theory side are perturbative in the large R-charge limit, so the original strong-weak AdS/CFT duality becomes precisely testable in this setting.   

The Penrose limit provides a new twist to the holography story. In the celebrated AdS/CFT holographic dictionary in \cite{Witten:1998qj}, the CFT lives at the boundary of a bulk AdS space and its local operators couple to the boundary configurations of the AdS bulk fields. However, although the pp-wave background (\ref{ppwave}) comes from a Penrose limit of the AdS space, the geometry is rather different. As such, it is not clear how to directly apply the standard AdS holographic dictionary, particularly in the situations with finite string interactions. Our approach in some previous papers \cite{Huang:2002wf, Huang:2002yt, Huang:2010ne, Huang:2019lso, Huang:2019uue} is to consider another corner of the parameter space in the BMN limit, focusing on the free gauge theory.  In this case, the string theory side becomes infinitely curved $\mu\sim \infty$, and strings are effectively infinitely long and tensionless, but can still have finite string interactions. Most interestingly, since the string spectrum is completely degenerate, the tensionless string can jump from one excited state to another without energy cost through a quantum unitary transition. It turns out that in this case the effective string coupling constant should be identified with a finite genus counting parameter $g:=\frac{J^2}{N}$, where $J$ is the large R-charge and scales like $J\sim \sqrt{N} \sim \infty$ in the BMN limit. Some higher genus BMN correlators were first computed in \cite{Kristjansen:2002bb,  Constable:2002hw}. 

Since the full fledged holographic dictionary is no longer available in the pp-wave background, our pragmatic approach is to try to compute the physical quantities on both sides of the correspondence and find potential non-trivial agreements. In this sense, a mismatch with naive expectation is not necessarily a contradiction of the holographic principle. Instead, one should focus on finding aspects where the calculations from both side do match, and try to give physical derivations or proofs of such mathematical coincidence. Besides the free string spectrum originally considered in \cite{Berenstein:2002jq}, some more tests of the pp-wave holography  are immediately clear. For example, the free planar  three-point functions of BMN operator should correspond to the Green-Schwarz light cone string field cubic vertex \cite{Green:1982tc, Green:1984fu} in the infinitely curved pp-wave background \cite{Huang:2002wf, Spradlin:2002ar}. In the papers \cite{Huang:2002yt, Huang:2010ne}, we further proposed a factorization formula, where the free higher genus BMN correlators are holographically related to string loop diagram calculations by pasting together the cubic string vertices without propagator. More recently, we propose a probability interpretation of the BMN two-point functions \cite{Huang:2019lso}. This also provides yet another interesting new entry of the pp-wave holographic dictionary that the BMN two-point function does not naively correspond to a quantum transition amplitude on the string theory side, but rather to its norm square. A consequence of the probability interpretation is the non-negativity of BMN two-point functions, which can be demonstrated for BMN operators with two stringy modes at any genus, or three stringy modes at genus one. In this paper, we further test the non-negativity conjecture for BMN operators with four and five stringy modes at genus one. Surprisingly, it turns out this is no longer valid. Of course, as mentioned earlier, this is not necessarily a contradiction of holographic principle according to our philosophy, but rather provide a new perspective on the limitation of our probability interpretation. 

Motivated by the results, we further check the factorization formula for the case of four stringy modes and confirm that it is still valid. We also check that the correspondence of planar three-point functions with Green-Schwarz string vertex is robust in the case of many string modes. Our mixed test results for this case shall motivate potential physical explanations which might shed new light on the still mysterious holographic principle. 

In some potentially related interesting recent developments, Gaberdiel and Gopakumar et al study string theory on a $AdS_3$ background, dual to a symmetric product CFT \cite{Eberhardt:2019ywk, Dei:2020zui, Gaberdiel:2020ycd}, with ideas dating back to some early papers e.g. \cite{Gopakumar:2003ns}. Although the technical details are rather different, there appears to be some common features with our works that the strings are tensionless and the dual CFT is free. To our knowledge, in various special situations where the higher genus string amplitudes can be systematically computed, our setting by far most resembles the usual critical superstring theory on flat spacetime, with of course still certain notable simplifications that in our case there is no continuous light cone or transverse momentum due to the infinite curvature and Ramond-Ramond flux in the background.  

The paper is organized as the followings.  In Sec. \ref{secreal} we review some notations and previous results, with an emphasis on the real and symmetric properties of the two-point functions. In Sec. \ref{seccalculations} we calculate the torus two-point functions of BMN operators with four string modes with the notations of some standard integrals. We also compute the case five string modes for the generic situations of mode numbers with no degeneracy. In both cases we discover that they are not alway non-negative. In Sec. \ref{secfactorization} we perform the one-loop string calculations and confirm that the factorization formula for the case of four string modes is still valid. In Sec. \ref{vertex} we check the correspondence of planar three-point functions with Green-Schwarz string vertex with many string modes. In Sec. \ref{secissues} we consider the situations of multiple string modes  in the same transverse direction. We conclude with some discussions in Sec. \ref{secdiscussion}.

\section{The reality of higher genus two-point functions with more stringy modes} \label{secreal}

Let us first introduce some notations for the higher genus two-point functions, and review some previous results. The integral formula is naively complex and we perform a more careful analysis of its reality property. The string vacuum state in the pp-wave geometry is described by a dual vacuum BMN operator with large R-charge $O^{J} = \Tr(Z^J)$, where $Z=\frac{1}{\sqrt{2}}(\phi^5+i\phi^6)$ is a complex scalar field in the $SU(N)$ adjoint representation, constructed from two of the six real scalar fields in the $\mathcal{N}=4$ $SU(N)$ super-Yang-Mills theory. We take the BMN limit $J\sim \sqrt{N}\sim \infty$ with $g:=\frac{J^2}{N}$ finite, and focus on free gauge theory. As in the previous papers, our notation omits the universal spacetime factors in the correlators. 

The stringy states with bosonic excited modes in the eight transverse directions are constructed by inserting the four remaining real scalars $\phi^I$ and four covariant derivatives $D_I$ where $I=1,2,3,4$ into the string of $Z$'s with phases. For example, the  BMN operators up to four scalar oscillator modes are the followings 
  \be \ba
   \label{BMNoperators}
& O^{J}  = \frac{1}{\sqrt{JN^J}}TrZ^J,  ~~~~
O^{J}_{0} = \frac{1}{\sqrt{N^{J+1}}} Tr(\phi^{I} Z^{J}),  \\
& O^J_{-m,m} = \frac1{\sqrt{JN^{J+2}}} \sum_{l=0}^{J-1}e^{\frac{2\pi iml}{J}} 
Tr(\phi^{I_1} Z^l\phi^{I_2} Z^{J-l}).  \\
& O^{J}_{(m_1,m_2,m_3)} = \frac{1}{\sqrt{N^{J+3}}J} \sum_{l_1, l_2=0}^{J}  e^{\frac{2\pi im_2l_1}{J}} e^{\frac{2\pi im_3l_2}{J}}  \Tr(\phi^{I_1} Z^{l_1} \phi^{I_2} Z^{l_2-l_1} \phi^{I_3} Z^{J-l_2}).  \\
& O^{J}_{(m_1,m_2,m_3, m_4)}  \\
&= \frac{1}{\sqrt{N^{J+4}}J^{\frac{3}{2}} } \sum_{l_1, l_2, l_3=0}^{J}  e^{\frac{2\pi im_2l_1}{J}} e^{\frac{2\pi im_3l_2}{J}}   e^{\frac{2\pi im_4l_3}{J}}   \Tr(\phi^1 Z^{l_1} \phi^2 Z^{l_2-l_1} \phi^3 Z^{l_3-l_2} \phi^4 Z^{J-l_3}). 
\ea \ee
Here one can use the cyclicity of the trace to move one scalar to the starting position for convenience, the mode numbers $\sum_i m_i=0$ in the case of three and four modes. The operators are properly normalized to be orthonormal at the genus zero or planar level. The convention is that the first operator $O^J$ corresponds to the closed string vacuum state, and the positive and negative modes in the other operators represent the left and right moving stringy excited modes, while the zero modes are supergravity modes representing discretized momenta in the corresponding traverse direction. The construction ensures only operators satisfying closed string level match conditioning are non-vanishing. As a consequence, the stringy excited states have at least two oscillator modes with opposite signs. Analogously, we can add more stringy modes and denote the properly normalized BMN operator $ O^J_{m_1,m_2,\cdots, m_k}$ with the closed string level matching condition $\sum_{i=1}^k m_k=0$. Unless otherwise specified, we use this notation to denote $k$ different string modes.

\begin{figure}
  \begin{center}
\includegraphics[width=6in]{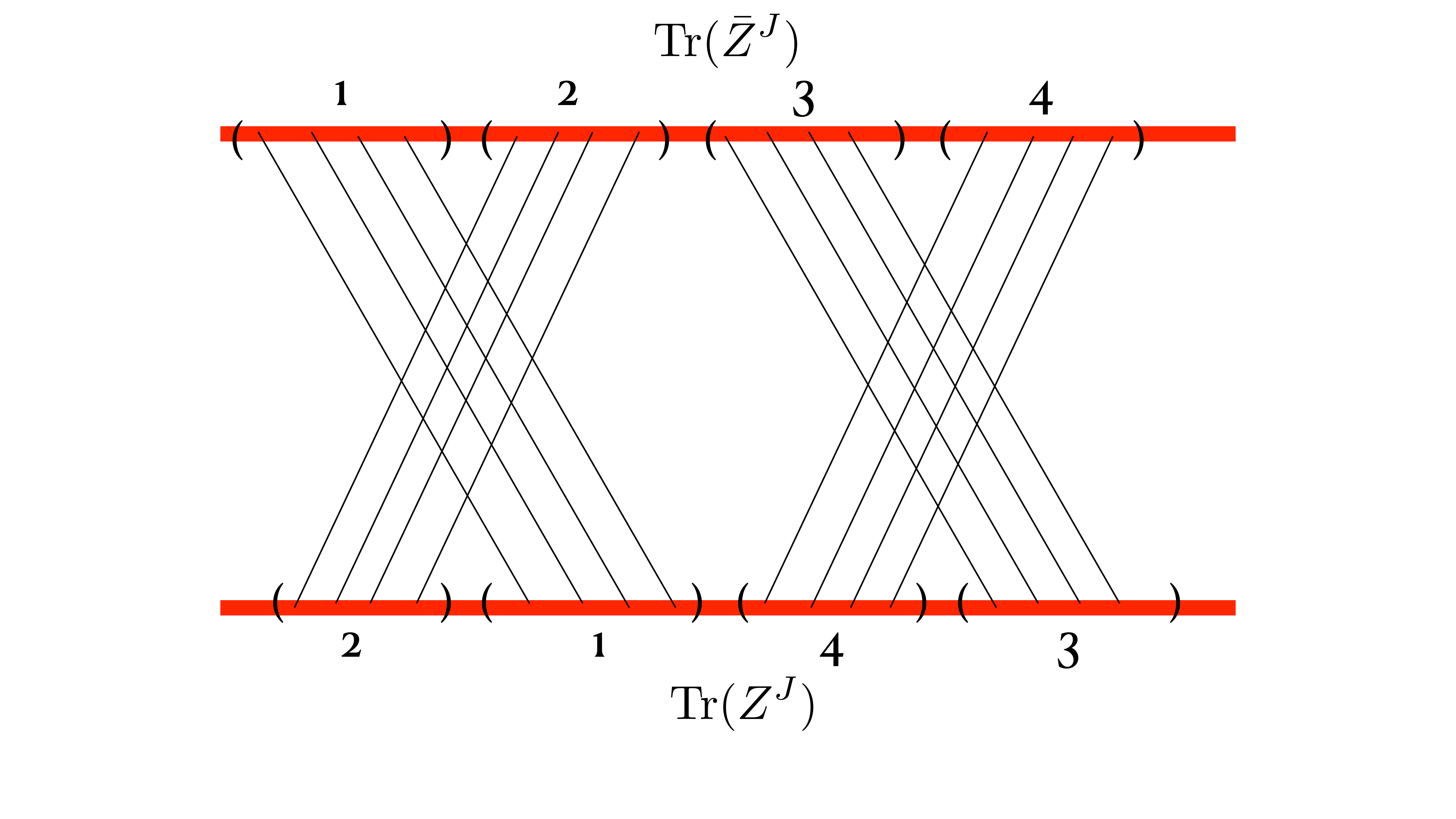} 
\end{center}
\caption{The torus diagram. }  \label{torusdiagram}
\end{figure}

The free two-point functions at higher genus $h\geq 1$ are computed by dividing the string of  $Z$'s up to $n\leq 4h$ segments and Wick contracted according to a permutation of $(1,2,\cdots ,n)$. We only consider cyclically inequivalent permutations where no two neighboring numbers are consecutive.  The contributions of such Feynman diagrams of genus $h$ are proportional to $\frac{J^{n}}{N^{2h}}$. So the dominant contributions come from those of the maximal number of segments $n=4h$ and we can neglect the other cases $n<4h$ which are suppressed in the large R-charge limit. Furthermore, in the BMN limit, the contributions are proportional to  $\frac{J^{4h}}{N^{2h}}= g^{2h}$, confirming the finite parameter $g$ as the genus counting parameter therefore the effective string coupling constant with our restriction to free gauge theory.  We should note that a generic permutation of $(1,2, \cdots, 4h)$ can give Feynman diagram with genus higher than $h$. A useful rule to select genus $h$ permutations is to generate them by string diagrams with $h$ loops \cite{Huang:2010ne}.  It is known that there are $\frac{(4h-1)!!}{2h+1}$ such genus $h$ permutations \cite{MR848681}. For example, at genus one there is only one such permutation, and can be generated by a one-loop string process $(1234)\rightarrow (12) (34) \rightarrow (2143)$.  The field theory torus diagram is depicted in Figure \ref{torusdiagram}. Please note that we denote the genus as $h$ because the usual symbol $g$ has been used as the effective string coupling. 

Once the string of $Z$'s is Wick contracted with $\bar{Z}$'s, we can add scalar insertions and contract them along the lines of $Z$'s to preserve the genus of the Feynman diagram. In the BMN limit each scalar insertion gives an integral with the corresponding phases. For example, the free torus two-point function can be written as 
\be \ba  \label{freetorus1}
& \bra \bar{O}^J_{(m_1,m_2,\cdots ,m_k)} O^J_{(n_1,n_2,\cdots ,n_k)} \ket_{\textrm{torus}} \\
&= \frac{g^2}{4} \int_0^1 dx_1dx_2dx_3dx_4 \delta(x_1+x_2+x_3+x_4-1) \times \\ 
&  \prod_{i=1}^{k}  (\int_0^{x_1} +e^{2\pi i n_i (x_3+x_4)}\int_{x_1}^{x_1+x_2}+e^{2\pi i n_i(x_4-x_2)}\int_{x_1+x_2}^{1-x_4}+e^{-2\pi i n_i(x_2+x_3)}\int_{1-x_4}^1) dy_i e^{2\pi i (n_i-m_i)y_i } \\
&= g^2 \int_0^1 dx_1dx_2dx_3dx_4 \delta(x_1+x_2+x_3+x_4-1) \int_0^{x_1} dy_k e^{2\pi i (n_k-m_k)y_k} \times \\ 
&  \prod_{i=1}^{k-1}  (\int_0^{x_1} +e^{2\pi i n_i (x_3+x_4)}\int_{x_1}^{x_1+x_2}+e^{2\pi i n_i(x_4-x_2)}\int_{x_1+x_2}^{1-x_4}+e^{-2\pi i n_i(x_2+x_3)}\int_{1-x_4}^1) dy_i e^{2\pi i (n_i-m_i)y_i },
\ea \ee
where in the second equality we use the cyclicity to put the one string mode into one of the four segments which have the same contribution since there is only one cyclically inequivalent permutation for the torus diagram. This breaks the symmetry between different modes but would be sometimes convenient for calculations. 

By definition the correlator is invariant under a complex conjugate and exchange of modes. We can perform a more careful analysis. We take the complex conjugate in the first formula in (\ref{freetorus1}), and change the integration variables $y_i\rightarrow 1-y_i, i=1,2,\cdots, k$ and ${x_1, x_2, x_3, x_4}\rightarrow {x_4, x_3, x_2, x_1}$.  After a simple calculation, also using the closed string level matching condition, one can check the formula remains the same. So the torus two-point function is purely real and symmetric.

The analysis for higher genus $h\geq 2$ is somewhat more complicated. When the $k$'s string mode runs through $4h$ segments as in the first equality in (\ref{freetorus1}), generically they correspond to multiple cyclically inequivalent permutations if we fix the $k$'s string mode in the first segment as in the second equality in (\ref{freetorus1}). To illustrate, we consider the case of genus two, which have 21 cyclically inequivalent permutations. These 21 permutations are divided into 4 groups
\be \ba \label{group2}
& 1.  ~ (14732865), (17548362), (18643725),  (14875326),  (15837642),  (18472653),  \\ 
& ~~~(15428736), (17625843), \\
& 2. ~ (15387426), (15842763), (16528473), (17362854), (17438625),  (14863275),  \\
&  ~~~ (16483752), (18537264), \\
& 3. ~ (14325876), (14765832),  (18365472),  (18725436), \\
& 4. ~ (16385274). 
\ea \ee
Here we have use the cyclicity to always put 1 into the first place in the permutations. Each group of permutations is generated by running a particular string mode through $8$ segments. A convenient rule is to start with a permutation, subtract each element by 1 (with 1 replaced by 8), then cyclically move 1 to the first position. One can repeat this operation until the original permutation reappears. The permutations of each group have the same multiplicity with respect to the string diagrams in the factorization formulas \cite{Huang:2010ne}. The contribution of each group to the genus 2 two-point function is real. However, the contribution of each individual permutation may be complex if we fix one particular, e.g. the $k$'s, string mode to be in the first segment as in the second equality in (\ref{freetorus1}). The permutations can then be further classified by some permutations whose contributions are real, and some other pairs where each pair consists of two permutations whose contributions are complex conjugate to each other. In our case, there are 5 self-conjugate permutations and 8 pairs of conjugate permutations as the followings 
\be \ba \label{pairs}
& {\rm Self ~conjugate}: ~~  (18365472), (14325876),  (14875326),  (17625843),  (16385274) , \\
& {\rm Conjugate ~pairs}:~~~   \{(1 4765832), (18725436) \},  ~~ \{(15837 642),  (18643725) \}, \\
&~~~~~~~~~~~~~~~~~~~~~~~~~~    \{ (16483752), (18537264) \}, \{ (17548362),  (184726 5 3) \},  \\
&~~~~~~~~~~~~~~~~~~~~~~~~~~   \{ (14863275),  (15387426) \}, \{(14732865), (15428736)\},  \\
&~~~~~~~~~~~~~~~~~~~~~~~~~~   \{(17438625),  (15842763) \},  \{(1 7 3628 5 4),  (1 6 5 2 8 4 7 3) \}. 
\ea \ee
The conjugate pairs always fall into the same group in (\ref{group2}). We can similarly perform the change of integration variables $y_i\rightarrow 1-y_i, i=1,2,\cdots, k$ and reverse the order of 8 segments as in the torus case to show that the two contributions of each pair in (\ref{pairs}) are complex conjugate to each other. This also provides a convenient rule for computing the conjugate permutation. One fixes 1 in the first position, then replace the 7 remaining numbers by $a\rightarrow 10-a$ and reverse their order, i.e. the conjugate of the permutation $(1,a_1,a_2,\cdots, a_7)$ is simply $(1,10-a_7, 10-a_6,\cdots, 10-a_1)$. 

Since a conjugate pair of permutations should have the same genus in our case, we expect this argument similarly works out at higher genus so the two-point functions are always real and symmetric
\be \ba
& \bra \bar{O}^J_{(m_1,m_2,\cdots ,m_k)} O^J_{(n_1,n_2,\cdots ,n_k)} \ket_h^* = \bra \bar{O}^J_{(m_1,m_2,\cdots ,m_k)} O^J_{(n_1,n_2,\cdots ,n_k)} \ket_h , \\
&\bra \bar{O}^J_{(m_1,m_2,\cdots ,m_k)} O^J_{(n_1,n_2,\cdots ,n_k)} \ket_h = \bra \bar{O}^J_{(n_1,n_2,\cdots ,n_k) } O^J_{(m_1,m_2,\cdots ,m_k) } \ket_h .
\ea \ee

An interesting formula, as discussed in \cite{Huang:2019lso}, is to sum over one set of string modes 
\be \label{normalization}
\sum_{\sum_{i=1}^k n_k=0 }\bra \bar{O}^J_{(m_1,m_2,\cdots ,m_k)} O^J_{(n_1,n_2,\cdots ,n_k)} \ket_{h} = 
\frac{(4h-1)!!}{(2h+1) (4h)!} g^{2h} . 
\ee
This can be also easily derived using the Poisson resummation formula $\sum_{n=-\infty}^{\infty} e^{2\pi i x n} = \sum_{p=-\infty}^{\infty}  \delta (x-p)$. The resulting integral after the summation can be easily performed due to delta function constrain, and is independent of the remaining set of string modes. One can thus sum up all genus contributions with a proper normalization by the all-genera formula of vacuum correlator
\be \label{matrixp}
p_{(m_1,m_2,\cdots ,m_k), (n_1,n_2,\cdots ,n_k)} = \frac{g}{2\sinh(\frac{g}{2})} \sum_{h=0}^{\infty} \bra \bar{O}^J_{(m_1,m_2,\cdots ,m_k)} O^J_{(n_1,n_2,\cdots ,n_k)} \ket_{h} . 
\ee
Then the matrix element is real and looks like a probability distribution 
\be \label{normalization1}
\sum_{\sum_{i=1}^k n_k=0 } p_{(m_1,m_2,\cdots ,m_k), (n_1,n_2,\cdots ,n_k)}  =  1.  
\ee

To interpret the matrix element (\ref{matrixp}) as a probability, it needs also to be non-negative. In order to keep the nice normalization relations (\ref{normalization}, \ref{normalization1}), we can not simply add some non-uniform phase factors to the BMN operators, so the signs of two-point functions can not be trivially changed and have physical relevance. Since the string coupling constant $g$ can be arbitrary, one may expect each correlator in the sum should be non-negative if such interpretation is valid. For the case of two string modes $k=2$, it can be easily shown that the correlators are indeed non-negative since the two string modes have opposite sign \cite{Huang:2019lso}. For the case of three string modes, one can also explicitly check that at the torus two-point function is always non-negative. For example, the torus two-point functions for generic case with no degeneracy in mode numbers is \cite{Huang:2010ne} 
\begin{eqnarray} \label{F14result}
 \bra \bar{O}^J_{(m_1,m_2,m_3)} O^J_{(n_1,n_2,n_3)} \ket_{\textrm{torus}}   
= \frac{g^2}{32\pi^4 }\frac{\sum_{i=1}^3(m_i-n_i)^2    }{\prod_{i=1}^3 (m_i-n_i)^2},
\end{eqnarray} 
which is manifestly positive. These properties strongly suggest a new entry of the pp-wave holographic dictionary 
\be \label{newentry} 
 p_{(m_1,m_2,\cdots ,m_k), (n_1,n_2,\cdots ,n_k)} = | \bra m_1,m_2,\cdots ,m_k | \hat{U} (g) |n_1,n_2,\cdots ,n_k\ket |^2, ~~~ k=2,3, 
\ee 
where the states $|n_1,n_2,\cdots ,n_k\ket $ denote the orthonormal BMN states of free string theory, while the operator $\hat{U} (g)$ describes the quantum unitary  transition between the tensionless strings. As discussed in \cite{Huang:2019lso}, the higher point functions are vanishing in the BMN limit and are regarded as virtual processes, so a single string can not actually decay into multi-strings through a finite physical process. In this sense the single strings form a complete Hilbert space  $\sum_{\sum_{i=1}^k n_k=0 }    |n_1,n_2,\cdots ,n_k\ket  \bra n_1,n_2,\cdots ,n_k| =I$. 

In the next Section \ref{seccalculations} we will calculate the cases of torus two-point functions with more than 3 string modes. It turns out the results are not always non-negative. Therefore we are currently restricting our probability interpretation and the proposed holographic dictionary (\ref{newentry}) to the situations with no more than 3 (different) string modes.

\section{The calculations of torus two-point functions}  \label{seccalculations}

In this section we provide some details of the calculations of the integral formula (\ref{freetorus1}) for $k=4,5$ string modes, generalizing previous results. First we introduce some standard integrals. For the case of $k=5$, we only compute a generic case, where no degeneracy of string modes appears in the standard integrals. It turns out that this is actually simpler than the case of four modes and we consider it first. We then study the case of $k=4$ modes in more details, and provide the universal result in terms of the standard integrals, which are valid for all cases including degeneracy.

\subsection{Some standard integrals}  \label{standardint}

The following standard integral, appeared in e.g. \cite{Constable:2002hw, Huang:2010ne},  is very useful for calculating higher genus correlators, 
\begin{eqnarray} \label{integral1}
I(u_1,u_2,\cdots,u_r) \equiv \int_0^1  dx_1\cdots dx_r \delta(x_1+\cdots+x_r-1) e^{ 2\pi i(u_1x_1+\cdots u_rx_r) }
\end{eqnarray} 
Here It is clear that the integral is unchanged if we add an integer to all the arguments. If some of the $u_i$'s are identical, one uses the following notation 
\begin{eqnarray} \label{combine}
I_{(a_1,\cdots,a_r)} (u_1,u_2,\cdots ,u_r)\equiv I(u_1,\cdots, u_1, u_2,\cdots ,u_2, \cdots ,u_r,\cdots ,u_r), 
\end{eqnarray}
where $a_i$'s are integers representing the numbers of the $u_i$'s in the right hand side, and for $a_i=0$ we can just eliminate the corresponding argument. The integral can be calculated by the following recursion relation  
\begin{eqnarray} \label{recursive}
&&2\pi i (u_i-u_j) I_{(a_1,\cdots,a_r)} (u_1,u_2,\cdots ,u_r)  \nonumber \\
&=& I_{(a_1,\cdots,a_j-1,\cdots, a_r)} (u_1,u_2,\cdots ,u_r) -I_{(a_1,\cdots,a_i-1,\cdots, a_r)} (u_1,u_2,\cdots ,u_r)  , 
\end{eqnarray}
If $u_i\neq u_j$ then this equation can be used to reduce the number of arguments, but the relation is also valid and both sides are zero when $u_i=u_j$. From the recursion relation one can obtain the formulas for the integral 
\begin{eqnarray} \label{integral3}
I(u_1,u_2,\cdots u_r) &=& \sum_{i=1}^r e^{2\pi iu_i} \prod_{j\neq i} \frac{1}{2\pi i(u_i-u_j)} , \\
 \label{integral4}
I_{(a_1+1,\cdots,a_r+1)}(u_1,\cdots,u_r) &=& \prod_{i=1}^r \frac{(\partial / \partial u_i)^{a_i}}{(2\pi i)^{a_i} a_i!} I(u_1,\cdots,u_r),
\end{eqnarray}
where the $u_i$'s are different. We note we have used the $i$ symbol for both the pure imaginary number and the product index, which are easy to distinguish and should not cause confusion. In our calculations, the arguments $u_i$'s will be always integers, so the exponential functions will be simply 1 in the end results.

\subsection{Five modes: the generic case} 

For the case of five distinct string modes, at least one of the modes is covariant derivative $D_I$. The structures of free field two-point functions are the same as the cases of scalar field insertions, as also implied by supersymmetry. So we can simply apply the integral formula (\ref{freetorus1}) regardless of the type of string modes. 

Using the reality of the integral (\ref{freetorus1}), it turns out the calculations are especially simple for the generic situation with no degeneracy. First we assume $m_i\neq n_i$ for all $i$'s. Then the $y_i$'s integrals can be performed 
\be \ba  \label{5mode}
& ~~ \bra \bar{O}^J_{(m_1,m_2,\cdots ,m_5)} O^J_{(n_1,n_2,\cdots ,n_5)} \ket_{\textrm{torus}} \\
&= \frac{g^2}{(2\pi i)^5 \prod_{i=1}^5 (n_i- m_i) }  \int_0^1 dx_1dx_2dx_3dx_4 \delta(x_1+x_2+x_3+x_4-1) (e^{2\pi i (n_5-m_5) x_1} -1) \\
&  \times \prod_{i=1}^{4}  [ e^{2\pi i (n_i-m_i) x_1} -1 + e^{-2\pi i m_i (x_1+x_2) } -  e^{2\pi i (-m_i x_1-n_i x_2) }
+e^{2\pi i (-n_i x_2+m_i x_4)}  \\ & -e^{2\pi i [(n_i-m_i)x_1-m_i x_2 +n_i x_4]} + e^{- 2\pi i n_2 (x_2+x_3)} -e^{2\pi i (n_2 x_1 +m_2 x_4)}  ]. 
\ea \ee
So this calculation becomes some standard 4-dimensional integrals. Due to the factor of $i^5$, we only need to compute the imaginary part of the integrals because of the reality of the result. This is quite simple for the  4-dimensional  case. For example, suppose the integers $u_i\neq u_j$ for any $i\neq j$, some results of the standard integrals are 
\be \ba \label{someresults}
& I(u_1, u_2, u_3, u_4)  =0, \\
 & I_{(2,1,1)} (u_1, u_2, u_3) = -\frac{1}{4\pi^2 (u_1-u_2) (u_1-u_3) } , \\
&  I_{(2,2)} (u_1, u_2)  = -\frac{1}{2\pi^2 (u_1-u_2)^2},   \\ 
& I_{(3,1)} (u_1, u_2) = \frac{1}{4\pi^2 (u_1-u_2)^2} -\frac{i}{4\pi (u_1-u_2)} , \\
&  I_{(4)} (u_1) = \frac{1}{6} . 
\ea \ee 
We see that most results are real and the only imaginary contribution appears in $ I_{(3,1)}$. Assuming no further degeneracy among the mode numbers, we can check that the only contributions to the $I_{(3,1)}$ type integral come from taking $l=1,2,3,4$ factor(s) $e^{2\pi i (n_i-m_i)}$ and the $5-l$ factor(s) of $-1$ in the integrand in (\ref{5mode}). The result is quite simple 
\be \ba  
& ~~ \bra \bar{O}^J_{(m_1,m_2,\cdots ,m_5)} O^J_{(n_1,n_2,\cdots ,n_5)} \ket_{\textrm{torus}} \\
&=  \frac{g^2}{(2\pi )^6 \prod_{i=1}^5 (n_i- m_i) } [ \sum_{i=1}^5 \frac{1}{n_i-m_i} - \sum_{i=1}^4 \sum_{j=i+1}^5  \frac{1}{n_i-m_i + n_j-m_j} ]. 
\ea \ee 
We can compute the result for some random mode numbers, and find that it can be either positive or negative.

The calculations actually work similarly for the cases of any odd number of stringy modes, providing a simpler method for obtaining the result for the case of three generic stringy modes with no degeneracy in (\ref{F14result}).

\subsection{Four modes} 
We use the second equality in (\ref{freetorus1}) which fixes the 4th string mode in the first segment, namely $0<y_4<x_1$. It turns out there are 20 cases where we can put the positions of remaining string modes $y_{1,2,3}$, up to some permutation symmetries. We write the 8-dimensional integrals in the standard form and list these 20 cases in the followings. 

\begin{enumerate}
\item The variables $0<y_1,y_2,y_3, y_4<x_1$.  The permutations of indices $1, 2, 3, 4$ give $4!=24$ such integrals.  Without loss of generality we consider $0<y_1<y_2<y_3<y_4<x_1$. After dissecting the integral, the contribution is 
\be
I_{(1,1,1,5)} (n_4-m_4, n_4-m_4+n_3-m_3, -n_1+m_1, 0). 
\ee

\item The variables $0<y_2<y_3<y_4<x_1<y_1<x_1+x_2$. There are $3!\cdot 3=18$ similar integrals by counting the choice of $y_1$ and permutations of indices $2,3,4$. The contribution is 
\be
I_{(2,2,2,1,1)} (-m_1, -n_1, 0, -m_1+n_4-m_4, -n_1-n_2+m_2).
\ee

\item The variables $0<y_2<y_3<y_4<x_1<x_1+x_2< y_1< x_1+x_2+x_3 $. There are also $3!\cdot 3=18$ similar integrals by counting the choice of $y_1$ and permutations of indices $2,3,4$. The contribution is 
\be
I_{(2,2,1,1,1,1)} (n_1-m_1, 0, -m_1, n_1, m_2-n_2,  n_1-m_1+n_4-m_4 ).
\ee

\item The variables $0<y_2<y_3<y_4<x_1< x_1+x_2+x_3<y_1<1 $. There are also $3!\cdot 3=18$ similar integrals by counting the choice of $y_1$ and permutations of indices $2,3,4$. The contribution is 
\be
I_{(2,2,2,1,1)} (m_1, n_1, 0,  n_1+n_4-m_4, m_1+m_2-n_2 ).
\ee

\item The variables $0<y_3<y_4<x_1<y_1<y_2<x_1+x_2 $. There are  $3\cdot 2 \cdot 2=12$ similar integrals by counting the choice of $y_3$ and exchange of indices between $3,4$ and between $1,2$. The contribution is 
\be
I_{(2,2,2,1,1)} (-m_1-m_2, -n_1-n_2, 0,  -n_1-m_2,  m_3+n_4 ). 
\ee

\item The variables $0<y_3<y_4<x_1<x_1+x_2<y_1<y_2<x_1+x_2+x_3 $. There are also $3\cdot 2 \cdot 2=12$ similar integrals by counting the choice of $y_3$ and exchange of indices between $3,4$ and between $1,2$. The contribution is 
\be
I_{(2,2,1,1,1,1)} ( n_1+n_2-m_1-m_2, 0 , n_1+n_2, -m_1-m_2, n_2-m_2, -n_3+m_3). 
\ee

\item The variables $0<y_3<y_4<x_1<x_1+x_2+x_3<y_1<y_2<1 $. There are also $3\cdot 2 \cdot 2=12$ similar integrals by counting the choice of $y_3$ and exchange of indices between $3,4$ and between $1,2$. The contribution is 
\be
I_{(2,2,2,1,1)} ( n_1+n_2, m_1+m_2, 0 , m_1+n_2, -n_3-m_4). 
\ee

\item The variables $0<y_3<y_4<x_1<y_1<x_1+x_2<y_2<x_1+x_2+x_3 $. There are also $3\cdot 2 \cdot 2=12$ similar integrals by counting the choice of $y_3$ and exchange of indices between $3,4$ and between $1,2$. The contribution is 
\be
I ( m_2, n_2, -m_1, -n_1, m_2+n_2, -m_1+n_2, m_2-n_1, m_2+n_2+m_3+n_4). 
\ee

\item The variables $0<y_3<y_4<x_1<y_1<x_1+x_2<x_1+x_2+x_3<y_2<1 $. There are also $3\cdot 2 \cdot 2=12$ similar integrals by counting the choice of $y_3$ and exchange of indices between $3,4$ and between $1,2$. The contribution is 
\be
I (m_2, n_2, -m_1, -n_1, 0, m_2-n_1, n_2-m_1, n_2-m_1+n_4-m_4 ). 
\ee

\item The variables $0<y_3<y_4<x_1<x_1+x_2<y_1<x_1+x_2+x_3<y_2<1 $. There are also $3\cdot 2 \cdot 2=12$ similar integrals by counting the choice of $y_3$ and exchange of indices between $3,4$ and between $1,2$. The contribution is 
\be
% I ( m_1, n_1, 0, m_1+m_2, n_1+n_2, -n_3-m_4, m_1+n_1+m_2, m_1+n_1+n_2). 
I ( m_2,  n_2 , -m_1, -n_1, -m_1-n_1, m_2-n_1, n_2-m_1, m_2+m_3+ n_2+n_4 ). 
\ee

\item The variables $0<y_4<x_1<y_1<y_2<y_3<x_1+x_2 $. There are $3!=6$ similar integrals by counting the permutations of indices $1,2,3$. The contribution is 
\be
I_{(2,2,2,1,1)}  ( m_4, n_4, 0, n_3+n_4-m_3 , m_1+m_4 -n_1). 
\ee

\item The variables $0<y_4<x_1<x_1+x_2<y_1<y_2<y_3<x_1+x_2+x_3 $. There are also $3!=6$ similar integrals by counting the permutations of indices $1,2,3$. The contribution is 
\be
I_{(2,2,1,1,1,1)}  ( n_4-m_4, 0, n_4, -m_4,-n_1+m_1, n_3+n_4-m_3-m_4). 
\ee

\item The variables $0<y_4<x_1<x_1+x_2+x_3<y_1<y_2<y_3<1 $. There are also $3!=6$ similar integrals by counting the permutations of indices $1,2,3$. The contribution is 
\be
I_{(2,2,2,1,1)}  ( -m_4, -n_4, 0, -m_4-m_3+n_3, -n_4+m_1-n_1). 
\ee

\item The variables $0<y_4<x_1<y_1<y_2<x_1+x_2<y_3<x_1+x_2+x_3 $. There are  also $3!=6$ similar integrals by counting the permutations of indices $1,2,3$. The contribution is 
\be
I  ( m_4, n_4, -m_3, -n_3, 0, m_4-n_3,  n_4-m_3, m_1+m_4+n_2+n_4). 
\ee

\item The variables $0<y_4<x_1<y_1<y_2<x_1+x_2<x_1+x_2+x_3<y_3<1 $. There are also $3!=6$ similar integrals by counting the permutations of indices  $1,2,3$. The contribution is 
\be
I  ( m_3, n_3, 0, m_3+m_4, n_3+n_4, -n_1-m_2, m_3+m_4+n_3). 
\ee

\item The variables $0<y_4<x_1<y_3<x_1+x_2< y_1<y_2<x_1+x_2+x_3 $. There are  also $3!=6$ similar integrals by counting the permutations of indices $1,2,3$. The contribution is 
\be
I  ( m_4, n_4, 0, m_3+m_4, n_3+n_4, -n_1-m_2, m_3+m_4+n_4, n_3+n_4+m_4). 
\ee

\item The variables $0<y_4<x_1<x_1+x_2< y_1<y_2<x_1+x_2+x_3 <y_3<1$. There are  also $3!=6$ similar integrals by counting the permutations of indices $1,2,3$. The contribution is 
\be
I  ( -m_4, -n_4, 0, m_1+m_2, n_1+n_2, n_2+m_1, m_1+m_2-n_4, n_1+n_2-m_4). 
\ee

\item The variables $0<y_4<x_1<y_3<x_1+x_2< x_1+x_2+x_3 <y_1<y_2<1$. There are  also $3!=6$ similar integrals by counting the permutations of indices $1,2,3$. The contribution is 
\be
I  ( -m_3, -n_3, 0, m_1+m_2, n_1+n_2, m_1+n_2, n_1+n_2-m_3, m_1+m_2-n_3). 
\ee

\item The variables $0<y_4<x_1<x_1+x_2<y_3< x_1+x_2+x_3 <y_1<y_2<1$. There are  also $3!=6$ similar integrals by counting the permutations of indices $1,2,3$. The contribution is 
\be
I  ( m_3, n_3, -m_4, -n_4, 0, n_3-m_4, m_3-n_4, m_1+m_3+n_2+n_3). 
\ee

\item The variables $0<y_4<x_1<y_1<x_1+x_2<y_2< x_1+x_2+x_3 <y_3<1$. There are  also $3!=6$ similar integrals by counting the permutations of indices $1,2,3$. The contribution is 
\be
I  ( m_2, n_2, -m_1, -n_1, m_2+m_3-n_1, m_2+m_3+n_2, n_2+n_3+m_2, n_2+n_3-m_1). 
\ee

\end{enumerate}

The 20 cases of integrals can be organized into 10 types of integrals, so that the total contribution to can be more succinctly written as
\be \ba  \label{4moderesult}
&~~ \bra \bar{O}^J_{(m_1,m_2, m_3, m_4)} O^J_{(n_1,n_2, n_3, n_4)} \ket_{{\rm torus}}  \\
& = g^2 \sum_{(i,j,k,l)} [ I_{(1,1,1,5)} (n_i-m_i, n_i-m_i+n_j-m_j, -n_k+m_k, 0)  \\
&+ I_{(2,2,2,1,1)} (-m_i, -n_i, 0, -m_i +n_j-m_j, -n_i-n_k+m_k)  \\
& +  I_{(2,2,2,1,1)} (m_i, n_i, 0, m_i - n_j + m_j,  n_i + n_k - m_k)  \\
& +  I_{(2,2,2,1,1)} (m_i+m_j, n_i+n_j, 0, m_i+n_j, -m_k-n_l )  \\
& +  I_{(2,2,1,1,1,1)} (n_i-m_i,0 , n_i, -m_i, -n_j+m_j, n_i-m_i +n_k-m_k)  \\
& + I(m_i, n_i, -m_j, -n_j, 0, m_i-n_j, n_i-m_j, n_i-m_j+n_k-m_k) \\
%& + I ( m_i, n_i, 0, m_i+m_j, n_i+n_j, m_i+n_i+m_j, m_i+n_i+n_j, -m_k-n_l) \\
& + I(m_i, n_i, -m_j, -n_j, -m_j-n_j, m_i-n_j, n_i-m_j, m_i+n_i+m_k+n_l)  \\
& + I(m_i, n_i, -m_j, -n_j, m_i+n_i, m_i-n_j, n_i-m_j, m_i+n_i+m_k+n_l)] \\
& + g^2  \sum_{(i,j) \leftrightarrow (k,l) }  I_{(2,2,1,1,1,1)} (n_i+n_j-m_i-m_j, 0, n_i+n_j, -m_i-m_j, n_i-m_i, -n_k+m_k) \\
& + g^2 \sum_{(i,j,k)}  I (m_i, n_i, -m_j, -n_j, m_i+n_i +m_k, m_i+n_i+n_k, m_i+m_k-n_j, n_i+n_k-m_j) .  \\
&\equiv g^2 \sum_{k=1}^{10} I_k
\ea \ee 
We provide some explanations of the notations. For later convenience we denote the 10 type of integrals by $I_k, k=1,2,\cdots , 10$, according to the order as written in the above equation, which should not be confused with labels of transverse space direction in the pp-wave geometry. The first 8 types of integrals are summed over the 24 permutations $(i,j,k,l)$ of $1234$. The notation $(i,j) \leftrightarrow (k,l)$ in $I_9$ denotes we sum only once if two permutations are related by exchanging $(i,j) \leftrightarrow (k,l)$. This can be achieved e.g. by specifying $1\in \{ i, k \} $ in the permutations. The last integral $I_{10}$ is summed over the 6 permutations  $(i,j,k)$ of $123$. Of the 10 types of integrals,  the $I_1$ comes from case 1 in the above enumeration, the $I_2$ from combining cases 2 and 13, the $I_3$ from combining cases 4 and 11, the $I_4$ from combing cases 5 and 7, the $I_5$ from combing cases 3 and 12, the $I_6$ from combining cases 9, 14 and 19, the $I_7$ from combing cases 10, 15 and 16, the $I_8$ from combining cases 8, 17 and 18, the $I_9$ from case 6, the $I_{10}$ from case 20. Although the last two integrals $I_9, I_{10}$ are not summed over the full 24 permutations of $1234$, it is easy to show they are also permutation symmetric using the closed string level matching conditions and the invariance of the standard integral under a shift of all arguments by an integer.  

We can perform the calculations using a compute program. The calculations are straightforward for a given set of mode numbers. An expression for the generic case where there is no further degeneracy in the arguments in (\ref{4moderesult}) can be obtained but it is too long to write down here. We can check some special cases. For example, when two modes $m_4=n_4=0$, this reduced to the case of three string modes considered in \cite{Huang:2010ne}. Another special case is when $m_i=0$ and $n_i\neq 0$, then the result identically vanishes, consistent with the conservation of discrete momentum in the transverse direction. 

The total contribution (\ref{4moderesult}) is always real, although each individual integral can be complex. Computing the results for some random mode numbers, we find that the result can be either positive or negative. We can provide some potentially helpful empirical observations about the signs of the torus two-point functions. However  for now there seems no particular strong motivation to warrant a thoroughly rigorous analysis. In the followings we assume all $m_i, n_i, i=1,2,3,4$ are non-zero. 

\begin{enumerate} 
\item If two pairs of mode numbers are the same, e.g. $m_i=n_i, i=1,2$, then the torus two-point functions are most likely positive. There may be some exceptions. For example, in the case $(m_i,n_i) =  (-10,-10), (-10,-10), (1,10), (19,10), i=1,2,3,4$, the torus two-point function is negative. If all mode numbers are the same, i.e. $m_i=n_i, i=1,2,3,4$, then we have not found an example of negative torus two-point function. 

\item For $m_i\neq n_i, i=1,2,3,4$, the sign of torus two-point function is most likely the same as $\prod_{i=1}^4 (m_i-n_i)$. There are also some exceptions. For example, in the case $(m_i,n_i) =  (8,5), (2,-6), (-5,-6), (-5,7) , i=1,2,3,4$, the torus two-point function is positive. This phenomenon can be explained from the previous method in the case of five modes in (\ref{5mode}). In the case of four modes we need to now pick up the real parts in the integrals (\ref{someresults}).  Only two terms give the last integral with completely degenerate arguments. So the result can be roughly written as 
\be \ba  
& ~~ \bra \bar{O}^J_{(m_1,m_2, m_3 ,m_4)} O^J_{(n_1,n_2, n_3 ,n_4)} \ket_{\textrm{torus}} \\
&= \frac{g^2}{(2\pi )^4 \prod_{i=1}^4 (m_i- n_i) }  (\frac{1}{3} + \cdots) , 
\ea \ee
where the $\cdots $ denotes some correction terms which are inverse squares of non-zero integers from mode numbers and also suppressed by a factor of $2\pi^2$, so their absolute values are most likely small comparing to $\frac{1}{3}$.

\end{enumerate}

\section{The factorization formula}  \label{secfactorization}

Since we have now discovered a new phenomenon in the case of more than three string modes that the BMN torus two-point functions can be negative, it is worthwhile to test the other proposals for the holographic dictionary, in particular the factorization formulas in \cite{Huang:2002yt, Huang:2010ne}. This also serves as a check of the somewhat complicated calculations in the previous Section \ref{seccalculations}. In this section we focus on the case of four string modes.

\subsection{Planar three-point functions} 

\begin{figure}
  \begin{center}
\includegraphics[width=6.5in]{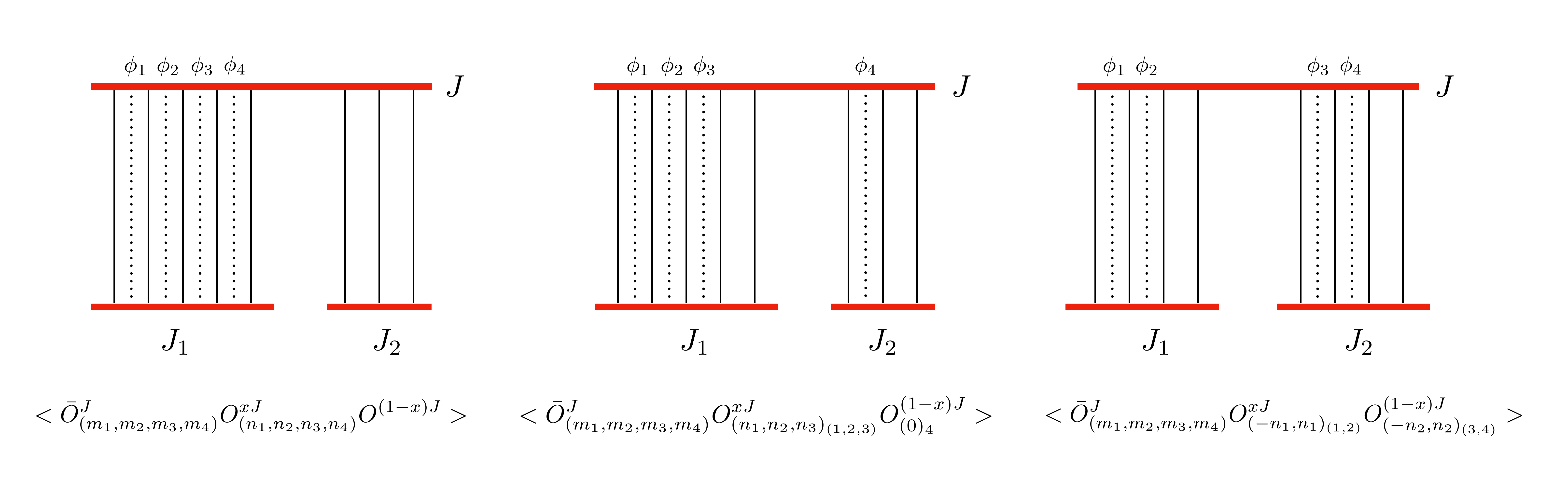} 
\end{center}
\caption{The planar three-point diagrams. }  \label{3pointdiagram}
\end{figure} 

First we shall calculate the relevant free planar three-point functions, which correspond to the  string vertices. There are 3 ways to distribute the 4 scalar insertions as the long string is cut into two short strings with $J_1\equiv xJ$ and $J_2\equiv (1-x)J$ number of $Z$'s ($0<x<1$). The field theory diagrams are depicted in Figure \ref{3pointdiagram}. We integrate over the positions of insertions with the BMN operators (\ref{BMNoperators}) and  compute the results
\be \ba \label{planar3}
& \bra \bar{O}^J_{m_1,m_2, m_3, m_4 }O^{J_1}_{k_1,k_2, k_3, k_4 } O^{ J_2} \ket \\ 
&= \frac{g}{\sqrt{J}} \frac{(1-x)^{\frac{1}{2}} }{x^{\frac{3}{2}}}  \int_{0}^x \prod_{i=1}^4 d y_i e^{-2\pi i (m_i-\frac{k_i}{x}) y_i }  
 =\frac{g}{\sqrt{J}} x^{\frac{5}{2}}   (1-x)^{\frac{1}{2}}  \prod_{i=1}^4 \frac{\sin(\pi m_i x) }{\pi (m_ix -k_i) } ,  \\
&\bra \bar{O}^J_{m_1,m_2, m_3, m_4 }O^{J_1}_{k_1,k_2, k_3 } O^{ J_2}_0 \ket  \\
&= \frac{g}{\sqrt{J} x} [ \int_{0}^x \prod_{i=1}^3 d y_i e^{-2\pi i (m_i-\frac{k_i}{x}) y_i } ][ \int_x^1 d y_4 e^{-2\pi i m_4 y_4} ]   = - \frac{g  x^2    }{\sqrt{J}}  [\prod_{i=1}^3 \frac{\sin(\pi m_i x) }{\pi (m_ix -k_i)}]   \frac{\sin(\pi m_4 x) }{\pi m_4}, \\
& \bra \bar{O}^J_{m_1,m_2, m_3, m_4 }O^{J_1}_{-k,k } O^{ J_2}_{-l,l}  \ket \\
&= g \frac{[ \int_{0}^x  d y_1dy_2  e^{-2\pi i (m_1+ \frac{k}{x}) y_1 } e^{-2\pi i (m_2 - \frac{k}{x}) y_2 } ]
  [ \int_x^1 dy_3 d y_4 e^{-2\pi i (m_3y_3+ \frac{l(y_3-x)}{1-x})  } e^{-2\pi i (m_4y_4 - \frac{l(y_4-x)}{1-x})  } ]   }{ \sqrt{J x(1-x)} } \\
& =  \frac{g}{\sqrt{J}}   [x(1-x)]^{\frac{3}{2}}  \frac{ \prod_{i=1}^4 \sin(\pi m_i x)  } {\pi^4 (m_1x+k) (m_2x-k) (m_3(1-x)+l) (m_4(1-x)-l) }.  
\ea \ee
For the simplicity of notation, we do not label the specific string modes in the operators with the implicit understanding that the string modes appearing in the same order in $\bar{O}$ and $O$ operators are the same. We note that the integral formulas are valid for any mode numbers, but the integrated results in the above equation may not be valid in some special cases where the denominator vanishes, e.g. some $m_i=k_i=0$. In those cases one needs to do the integral separately. As in the cases with less stringy modes, the three-point functions are always suppressed by a factor $\sqrt{J}$, so they are vanishing or ``virtual" in the BMN limit $J\sim \infty$.

\subsection{One-loop string diagram calculations}

\begin{figure}
  \begin{center}
\includegraphics[width=6in]{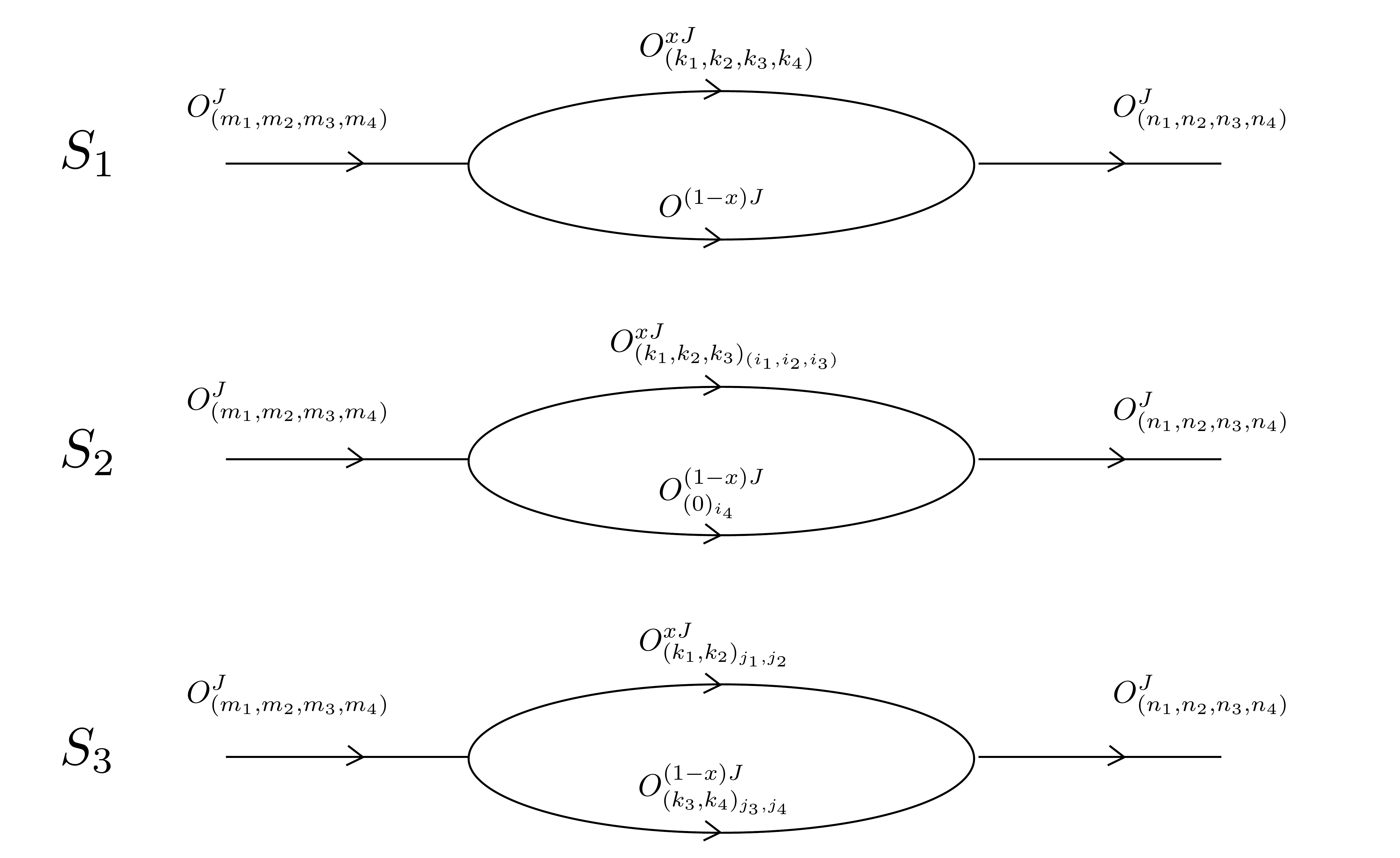} 
\end{center}
\caption{The one-loop string diagrams. }  \label{oneloop}
\end{figure}

There are also 3 string one-loop processes corresponding to the torus two-point functions, depicted in Figure \ref{oneloop}. In the BMN limit, the sum over the operator length becomes integral $\sum_{J_1=1}^{J-1} = J\int_0^1 dx $. We denote the contributions $S_1, S_2, S_3$ as the followings 
\be \ba \label{stringloop}
& S_1 = J \int_0^1 dx \sum_{\sum_i k_i=0}  \bra \bar{O}^J_{m_1,m_2, m_3, m_4 }O^{J_1}_{k_1,k_2, k_3, k_4 } O^{ J_2} \ket  
\bra \bar{O}^{J_1}_{k_1,k_2, k_3, k_4 } \bar{O}^{ J_2}  O^J_{n_1,n_2, n_3, n_4 } \ket  , \\
& S_2 =  J \int_0^1 dx \sum_{i_4=1}^4 \sum_{\sum_i k_i=0}   \bra \bar{O}^J_{m_{i_1}, m_{i_2}, m_{i_3}, m_{i_4} }O^{J_1}_{k_1,k_2, k_3 } O^{ J_2}_0  \ket  \bra \bar{O}^{J_1}_{k_1,k_2, k_3 } \bar{O}^{ J_2}_0  O^J_{n_{i_1}, n_{i_2}, n_{i_3}, n_{i_4} } \ket, \\
& S_3 =  J \int_0^1 dx \sum_{i_2=2}^4 \sum_{k,l =-\infty}^{+\infty}   \bra \bar{O}^J_{m_1,m_{i_2}, m_{i_3}, m_{i_4}  }O^{J_1}_{-k,k } O^{ J_2}_{-l,l}  \ket  \bra \bar{O}^{J_1}_{-k,k } \bar{O}^{ J_2}_{-l,l}   O^J_{n_1,n_{i_2}, n_{i_3}, n_{i_4}  } \ket ,
\ea \ee
where $(i_1, i_2, i_3, i_4)$ in $S_2$  is a cyclic permutation of $(1234)$ and $(i_2, i_3, i_4)$ in $S_3$ is a cyclic permutation of $(234)$. 

There are two methods for the computations of the equations in (\ref{stringloop}). The first method is to directly sum over the integrated results in (\ref{planar3}), then perform the $x$ integral. One need to use some summation formulas, which is mostly straightforward. But this method needs to deal with degenerate special cases separately.  The second method, discussed in \cite{Huang:2010ne} for the case of two string modes, is to directly use the integral formulas for planar three-point functions and sum over the intermediate modes first. This can be done using the Poisson summation formula. The resulting integrals with delta function constrains can then be converted to the standard integrals in Sec. \ref{standardint}. The second method works universally with any string mode numbers without the need to deal with degenerate cases separately, but the careful dissections of integral domains are also quite complicated.   

Since the sums and integrals here are always convergent, the two methods are equivalent.  The first method was used in the case of three string modes in \cite{Huang:2010ne}. However, in our current case of four string modes, the calculations become too complicated to obtain analytic answers in some steps. In any case, the explicit results are quite long and complicated to provide useful physical insight, so it is better to use the second method and write the contributions as the standard integrals in Sec. \ref{standardint}. We also use the first method complementarily to perform some numerical tests of the results. 

Summing over the intermediate string modes, we find the integral formulas for string loop diagram contributions 
\be \ba \label{Sformulas}
& S_1 = g^2 \int_0^1 (1-x) dx [\int_0^x \prod_{i=1}^4 dy_i dy^{\prime}_i e^{2\pi i (n_iy^{\prime}_i-m_iy_i )} ] [\prod_{i=1}^3 \sum_{p_i=-\infty}^{+\infty}  \delta(y_i^{\prime} -y_i +y_4 -y^{\prime}_4 -p_i x) ] , \\
& S_2 = g^2 \int_0^1 dx  [\int_0^x \prod_{i=1}^3 dy_i dy^{\prime}_i ] [\int_x^1 d y_4 d y^{\prime}_4]  [\prod_{i=1}^2 \sum_{p_i=-\infty}^{+\infty}  \delta(y_i^{\prime} -y_i +y_3 -y^{\prime}_3 -p_i x) ]   \\
& ~~~~~\cdot e^{2\pi i \sum_{i=1}^4 ( n_iy^{\prime}_i-m_iy_i )}   + \sum_{i=1}^3 (i\leftrightarrow 4) ,  \\
& S_3 = g^2 \int_0^1 dx  [\int_0^x \prod_{i=1}^2 dy_i dy^{\prime}_i ] [\int_x^1 \prod_{i=3}^4 d y_i d y^{\prime}_i]  [ \sum_{p_1, p_2 =-\infty}^{+\infty} \delta(y_1^{\prime} -y_1 +y_2 -y^{\prime}_2 -p_1 x)   \\  & ~~~~~~\cdot  \delta(y_3^{\prime} -y_3 +y_4 -y^{\prime}_4 -p_2 (1-x)) ] e^{2\pi i \sum_{i=1}^4 ( n_iy^{\prime}_i-m_iy_i )}    +\sum_{i=3}^4  (i\leftrightarrow 2), 
\ea \ee
where the contributions in $S_2, S_3$ have some extra terms which are simply related to the explicit expression by permutations of indices. After carefully dissecting the integral formulas, with details explained in the Appendix \ref{details}, we find that the contributions can be transformed into the 10 types of integrals in (\ref{4moderesult}). Specifically, the results are 
\be \ba 
& S_1 = g^2(2I_1 + I_2 + I_3 +I_4), \\
& S_2 = g^2(I_2 +I_3+ 2I_5 +2I_6 +I_7 +I_8), \\
& S_3 = g^2(I_4 + +I_7 +I_8 +2I_9 +2I_{10} ). 
\ea \ee
So we confirm an entry of the holographic dictionary of factorization formulas 
\be \label{dictionary2}
2 \bra \bar{O}^J_{(m_1,m_2, m_3, m_4)} O^J_{(n_1,n_2, n_3, n_4)} \ket_{{\rm torus}}  =S_1+S_2+S_3 =2g^2\sum_{i=1}^{10}I_i  .
\ee

\section{Comparison with light cone string field cubic vertex} \label{vertex} 

In this section we compare the planar three-point functions with many string modes with the Green-Schwarz light cone string field cubic vertex, generalizing the earlier paper for the case of two string modes \cite{Huang:2002wf}. The bosonic part of the Green-Schwarz cubic string field vertex can be described by $|V\ket =E_a|0\ket$, where the operator 
\be  \label{expo}
E_a\sim \exp [ \sum_{r=1}^2 \sum_{I=1}^8 \sum_{m,n=-\infty}^{\infty} a_{m(3)}^{I \dagger } N^{3,r}_{m,n} a_{n(r)}^{I \dagger} ] . 
\ee
We provide some explanations of the notations. The indices $r=1,2,3$ labels the three strings, with the convention that $r=3$ string has the largest light cone width which is the sum of those of $r=1,2$.  The bosonic operator $a^{I \dagger}_{m(r)}$ creates the $r$'s string state in the $I$'s transverse direction with BMN mode number $m$. The Neumann matrix encodes the string interactions. Its element $N^{3,3}_{m,n}=0$ for any $m,n$, and it has a symmetry $N^{r,s}_{m,n} = N^{s,r}_{n, m}$. Since the number of string modes in the 3rd long string is the sum of those of the two $r=1,2$ short strings, we only need to include  the $N^{3,r}_{m,n} $ type of matrix elements. This corresponds to the calculations of free planar BMN three-point functions  where the string modes are contracted between a long string and the two short strings.

The Neumann matrix elements were computed in the pp-wave background \cite{Spradlin:2002ar} and becomes much simplified in the infinite curvature limit \cite{Huang:2002wf}. We denote the light cone width of the two short strings as $x$ and $1-x$, corresponding to the relative lengths of the two short operators in the free planar three-point function. The relevant matrix elements in the infinite curvature limit are 
\be \ba \label{Neumann}
&N^{3,1}_{0,0} = \sqrt{x} , ~~~~  N^{3,2}_{0,0} = \sqrt{1-x}, \\
& N^{3,1}_{m,n}  = \frac{\sqrt{x} \sin(\pi m x)} {\pi (mx-n)}, ~~~~ N^{3,2}_{m,n}  = -\frac{\sqrt{1-x} \sin(\pi m x)} {\pi [m(1-x)-n]}, ~~~~ {\rm for }~(m,n)\neq (0,0), 
\ea \ee
We should note that we use a different convention for the basis of the bosonic creation operators from the literature \cite{Spradlin:2002ar, Huang:2002wf}. Due to the different conventions, there are also some sign differences in the Neumann matrix elements with the literature. The current convention is most convenient from the field theory perspective.   

In the study of superstring field theories in flat space \cite{Green:1982tc, Green:1984fu}, besides the cubic vertex, there are other important physical quantities, such as the prefactor and the higher order contact interactions. This is further studied in the pp-wave background and the dual field theory in e.g. \cite{Pankiewicz:2002tg, Gursoy:2002fj}. With our specialization to the infinite curvature limit, the tensionless strings do not have an effective action description. So in our case it seems that at tree level, the cubic vertex  $|V\ket $ is the only remaining relevant finite physical object to consider.

Suppose three BMN operators $O_1, O_2, O_3$ correspond to three string states $|1\ket, |2\ket, |3\ket$, then the planar three-point functions are related to the string vertex 
\be  \label{correspondence}
\frac{\bra \bar{O}_3 O_1 O_2 \ket }{\bra \bar{O}^J O^{xJ} O^{(1-x)J} \ket}  =  \frac{\bra 1 |\bra 2| \bra 3| V\ket } {\bra 0|V\ket} , 
\ee
where $|0 \ket $ is the string vacuum state, and the normalization factor of BMN vacuum correlator is simply 
\be
\bra \bar{O}^J O^{xJ} O^{(1-x)J} \ket =  \frac{g \sqrt{x(1-x)}}{\sqrt{J}}   .
\ee
The right hand side of (\ref{correspondence}) can be computed by expanding the bosonic operator (\ref{expo}) to appropriate order and extract the relevant Neumann matrix elements. For example, for the case of four string modes, the BMN operators correspond to the string states as 
\be \ba
& O^J_{m_1,m_2, m_3, m_4 } \Longleftrightarrow a^{I_1\dagger}_{m_1 (3)}  a^{I_2\dagger}_{m_2 (3)}  a^{I_3\dagger}_{m_3 (3)}  a^{I_4\dagger}_{m_4 (3)}  |0\ket , \\
& O^{J_1}_{k_1,k_2, k_3, k_4 } \Longleftrightarrow a^{I_1\dagger}_{k_1 (1)}  a^{I_2\dagger}_{k_2 (1)}  a^{I_3\dagger}_{k_3 (1)}  a^{I_4\dagger}_{k_4 (1)}  |0\ket, \\
& O^{J_1}_{k_1,k_2, k_3 } \Longleftrightarrow a^{I_1\dagger}_{k_1 (1)}  a^{I_2\dagger}_{k_2 (1)}  a^{I_3\dagger}_{k_3 (1)}   |0\ket, \\ 
& O^{J_1}_{-k, k } \Longleftrightarrow a^{I_1\dagger}_{-k (1)}  a^{I_2\dagger}_{k (1)}  |0\ket, ~~~ 
O^{J_2}_{-l, l } \Longleftrightarrow a^{I_3\dagger}_{-l (2)}  a^{I_4\dagger}_{l (2)}  |0\ket, \\
&O^{ J_2} \Longleftrightarrow  |0\ket  , ~~~ O^{ J_2}_0 \Longleftrightarrow a^{I_4\dagger}_{0 (2)}   |0\ket. 
\ea \ee
We can expand the exponential operator (\ref{expo}) to 4th order and compute planar three-point functions with the usual commutation relation of creation and  annihilation operators. The only non-vanishing contributions come from the 4th order which provides the same numbers of creation operators as those of the annihilation operators. The results are  
\be \ba 
& \bra \bar{O}^J_{m_1,m_2, m_3, m_4 }O^{J_1}_{k_1,k_2, k_3, k_4 } O^{ J_2} \ket 
=\frac{g \sqrt{x(1-x)}}{\sqrt{J}}  \prod_{i=1}^4 N^{3,1}_{m_i,k_i}  ,  \\
&\bra \bar{O}^J_{m_1,m_2, m_3, m_4 }O^{J_1}_{k_1,k_2, k_3 } O^{ J_2}_0 \ket  
=\frac{g \sqrt{x(1-x)}}{\sqrt{J}}    (\prod_{i=1}^3 N^{3,1}_{m_i,k_i}) N^{3,2}_{m_4, 0},  \\
& \bra \bar{O}^J_{m_1,m_2, m_3, m_4 }O^{J_1}_{-k,k } O^{ J_2}_{-l,l}  \ket 
=\frac{g \sqrt{x(1-x)}}{\sqrt{J}}   N^{3,1}_{m_1, -k} N^{3,1}_{m_2, k} N^{3,2}_{m_3, -l} N^{3,2}_{m_4, l}
.  
\ea \ee
This agrees with the field theory results (\ref{planar3}) using the Neumann matrix elements in (\ref{Neumann}). One can also check the case of three string modes previously computed in \cite{Huang:2010ne} and various degenerate cases. Of course, as we mentioned, the planar three-point functions are vanishing in the BMN limit $J\sim \infty$ and regarded as virtual processes, but their ratios with the vacuum correlator are finite and meaningfully related to the Neumann matrix elements.

It is not difficult to see that the Neumann matrix elements in (\ref{Neumann}) simply correspond to the integrations of the positions of relevant string mode with phases in the BMN operators with proper normalization, using the closed string level matching condition in the long 3rd string to cancel out an overall phase. We infer that although the physical setting has at most eight string modes of distinct directions, the mathematical structure of the holographic dictionary (\ref{correspondence}) is quite robust and survives even in a hypothetical situation with any number of different string modes, i.e. not just valid for $I=1,2, \cdots 8$. 

The analysis here also provides another perspective on the reality condition of higher genus two-point functions discussed in Sec. \ref{secreal}. Since the Neumann matrix elements are all real, the planar three-point functions are also always real. If the factorization formulas e.g. (\ref{dictionary2}) are correct, the higher genus two-point functions can be computed from string diagrams and must be real.

\section{Some issues with multiple string modes in the same transverse direction}  \label{secissues} 

Since we have now studied BMN operators with many string modes, it is appropriate to consider the situation of multiple modes in the same transverse direction. To our knowledge, this situation has not been much discussed in the literature. Naively, the corresponding BMN operators can be similarly constructed, using the same scalar field (or covariant derivative) going through the string of $Z$'s with multiple sums with phases, with possibly a different normalization discussed below. 

For simplicity we consider BMN operators with only the 4 scalar field insertions. First we introduce some notations, if multiple string modes correspond to the same direction, we use a square bracket to enclose the mode numbers. For example, the BMN operator with two identical scalar fields is denoted $O^J_{[-m,m]}$, and the BMN operator with three string modes where two of them have the same direction is denoted $O^J_{([m_1, m_2], m_3)}$. The closed string level matching condition is still the same that all mode numbers should sum to zero.   Since the scalar fields in the square bracket are exchangeable, e.g. the operators $O^J_{([m_1, m_2], m_3)}$ and $O^J_{([m_2, m_1], m_3)}$ are the same, we can choose to order the mode numbers in the square bracket, e.g. in a non-decreasing order.

However, this brings a subtle issue. We recall that the chiral primary operators with lowest dimension in a short multiplet of $\mathcal{N}=4$ super-Yang-Mills theory are constructed by the 6 real scalars in the $SO(6)$ symmetric traceless representation, see e.g. the review \cite{Aharony:1999ti, DHoker:2002nbb}. They are BPS operators whose conformal dimensions are protected by supersymmetry. When a real scalar appears multiple times, an operator may no longer be chiral primary. For example, the operator $\Tr((\phi^I)^2)$, known as the Konishi operator, is not a chiral primary operator, since it is not traceless in the $SO(6)$. The conformal dimension of this operator would grow at least as $(g_{YM}^2 N)^{\frac{1}{4}}$. On the other hand, the BMN vacuum operator $\Tr(Z^J)$ is a chiral primary operator since a power of the complex scalar $Z$ is automatically traceless in the $SO(6)$.

In the original calculations of planar anomalous conformal dimensions of the BMN operator  $O^J_{-m,m}$ \cite{Berenstein:2002jq}, one used the fact that for $m=0$, the operator $O^J_{0,0}$ is a chiral primary operator whose conformal dimension is not corrected by gauge interactions. So one only needs to compute the mode number $m$-dependent part which is perturbative in an effective gauge coupling constant $\lambda^{\prime} \equiv \frac{g_{YM}^2 N}{J^2} $, a small parameter in the BMN limit. In this sense the BMN operators of distinct scalar field insertions with non-zero modes are ``near BPS" operators. As mentioned, if we put two identical real scalars into the string of $Z$'s, the zero mode operator, namely $O^J_{[0,0]}$, is no longer a chiral primary operator. There may be large (field theory) quantum corrections to the $m$-independent part of its conformal dimension. So in this case the calculations of planar conformal dimension is no longer reliable. We are not aware a simple natural fix which also matches the expectations from the string theory side. 

In any case, we may hope by restricting ourselves to free gauge theory, this issue with large quantum gauge corrections does not cause problems.  We shall retest our earlier results for the cases involving BMN operators with multiple identical scalar fields. We find that the comparison with Green-Schwarz light cone string field cubic vertex \cite{Huang:2002wf} and factorization formula \cite{Huang:2002yt, Huang:2010ne} still go through smoothly. However, the probability interpretation \cite{Huang:2019lso} begins to encounter an issue in the case of three scalar field insertions with two of them identical. 

First we consider the case of two identical scalar fields. The BMN operators are 
\be \ba \label{2samemodes}
& O^J_{[-m,m]} = \frac1{\sqrt{JN^{J+2}}} \sum_{l=0}^{J-1}e^{\frac{2\pi iml}{J}} 
Tr(\phi^{I} Z^l\phi^{I} Z^{J-l}), ~~~~ m>0, \\
& O^J_{[0, 0]} = \frac1{\sqrt{2JN^{J+2}}} \sum_{l=0}^{J-1}
Tr(\phi^{I} Z^l\phi^{I} Z^{J-l}), 
\ea \ee
where $\phi^I$ is any one of 4 remaining real scalars. We only need to consider $m\geq 0$ since the negative $m$ gives the same operator. The zero mode has an extra normalization factor $\sqrt{2}$ to keep them orthonormal at the planar level. 

We consider the comparison with string field vertex in Sec. \ref{vertex}.  As an example we compute the vertex amplitude with three string states  $a^{I_1 \dagger} _{0(1)} |0\ket,   a^{I_2 \dagger} _{ 0(2)} |0\ket, a^{I_1 \dagger} _{-m(3)}  a^{I_2 \dagger} _{m(3)} |0\ket$. 
We have an extra contribution if the directions are the same $I_1=I_2$, namely,
\be
\frac{ \bra 0 | a^{I_1 \dagger} _{0(1)}  a^{I_2 \dagger} _{ 0(2)}  a^{I_1 } _{-m(3)}  a^{I_2 } _{m(3)}  |V\ket } { \bra 0  |V\ket}  =
\begin{cases}
   N^{3,1}_{-m, 0}  N^{3,2}_{m, 0}  ,      &  I_1\neq I_2  \\
   N^{3,1}_{-m, 0}  N^{3,2}_{m, 0}  +  N^{3,1}_{m, 0}  N^{3,2}_{-m, 0} ,       & I_1=I_2  .
\end{cases} 
\ee
The extra contribution for $I_1=I_2$ also appears in the extra contraction for identical scalar fields in the field theory calculations. So the comparison of BMN three-point functions with cubic string vertex is still valid in the case of multiple modes in the same direction. 

The factorization formula also works in this case. We note that with the extra contraction due to identical scalars, for $m,n\neq 0$ we have the formula
\be \ba
&  \bra \bar{O} ^J_{[-m,m]} O^{J_1}_0 O^{J_2}_0 \ket = 2 \bra \bar{O} ^J_{-m,m} O^{J_1}_0 O^{J_2}_0 \ket, \\ 
&  \bra \bar{O} ^J_{[-m,m]} O^{J_1}_{[-n,n]}  O^{J_2}  \ket  = \bra \bar{O} ^J_{-m,m} O^{J_1}_{-n,n}  O^{J_2}  \ket + \bra \bar{O} ^J_{-m,m} O^{J_1}_{n, -n}  O^{J_2}  \ket ,  \\ 
& \bra \bar{O} ^J_{[-m,m]}  O^J_{[-n,n]} \ket_{\rm torus} = \bra \bar{O} ^J_{-m,m}  O^J_{-n,n} \ket_{\rm torus} 
+ \bra \bar{O} ^J_{-m,m}  O^J_{n, -n} \ket_{\rm torus} ,  
\ea \ee
where $J=J_1+J_2$ and the three-point functions without label are planar. Using the fact $\bra \bar{O} ^J_{-m,m} O^{J_1}_0 O^{J_2}_0 \ket = \bra \bar{O} ^J_{m,-m} O^{J_1}_0 O^{J_2}_0 \ket $, $ \bra \bar{O} ^J_{-m,m} O^{J_1}_{-n,n}  O^{J_2}  \ket  = \bra \bar{O} ^J_{m, -m} O^{J_1}_{n, -n}  O^{J_2}  \ket $ and the factorization formula for the case of two different modes \cite{Huang:2010ne, Huang:2019lso}, we can write the analogous formula for the current case 
\be \ba
2 \bra \bar{O} ^J_{[-m,m]}  O^J_{[-n,n]} \ket_{\rm torus}  &= \sum_{J_1=1}^{J-1} \sum_{k=0}^{\infty} 
 \bra \bar{O} ^J_{[-m,m]} O^{J_1}_{[-k,k] }  O^{J_2}  \ket  \bra   \bar{O}^{J_1}_{[-k,k]}  \bar{O}^{J_2} O ^J_{[-n,n]}  \ket  \\
 &+  \sum_{J_1=1}^{[\frac{J}{2}] }  \bra \bar{O} ^J_{[-m,m]} O^{J_1}_{0 }  O^{J_2}_0  \ket  \bra   \bar{O}^{J_1}_{0}  \bar{O}^{J_2}_0 O ^J_{[-n,n]}  \ket. 
\ea  \ee 
We note that the difference is that we only need to sum over $k\geq 0$ and the second sum is over $J_1\leq J_2$ since the scalars in the two operators $O^{J_1}_0$ and $O^{J_2}_0$ are the same. The formula for case of $m,n=0$ is much simpler and also works in this case, taking into account the normalization in (\ref{2samemodes}). 

The calculations with more stringy modes are similar based on the experience. So we conclude that as long as the comparison with cubic string vertex and the factorization formula are valid in the case of many different string modes, then identifying some of the modes as the same shall not cause problems. 

Now we consider the probability interpretation for two and three string modes. Identifying some modes to the same apparently does not change the non-negativity of the correlators. So we only need to consider the normalization relation. For the case two string modes in the same direction we still also have the normalization relation similar as (\ref{normalization}) 
\be
\sum_{n=0 }^{\infty} \bra \bar{O}^J_{[-m,m] } O^J_{ [-n, n])} \ket_{h} = 
\frac{(4h-1)!!}{(2h+1) (4h)!} g^{2h} , 
\ee
where we now only need to sum over non-negative integer $k$. The formula is valid for both $m=0$ and $m>0$ since the zero mode decouple from non-zero modes. 

However, the normalization relation encounters a problem in the case of three string modes with two of them in the same direction. The BMN operators are the followings  
\be \ba
 O^{J}_{([m_1,m_2],m_3)} = 
 \frac{c}{\sqrt{N^{J+3}}J} \sum_{l_1, l_2=0}^{J}  e^{\frac{2\pi im_2l_1}{J}} e^{\frac{2\pi im_3l_2}{J}}  \Tr(\phi^1 Z^{l_1} \phi^1 Z^{l_2-l_1} \phi^2 Z^{J-l_2}). 
\ea \ee
Comparing to the case of three different modes (\ref{BMNoperators}), we add a normalization constant which is $c=1$ if $m_1<m_2$ and $c=\frac{1}{\sqrt{2}}$ if $m_1=m_2$, so that the operators are orthonormal at the planar level. Again we compute the sum over one set of mode numbers. Suppose  $m_1<m_2$, 
\be \ba
&~~ \sum_{n_1\leq n_2 } \bra \bar{O}^J_{( [m_1,m_2],m_3) } O^J_{( [n_1, n_2],n_3 )} \ket_{h} \\
&=  \sum_{n_1\neq  n_2 } \bra \bar{O}^J_{( m_1,m_2 ,m_3) } O^J_{( n_1, n_2 ,n_3 )} \ket_{h}  
+\sqrt{2} \sum_{n } \bra \bar{O}^J_{( m_1,m_2 ,m_3) } O^J_{( n, n ,n_3 )} \ket_{h}  . 
\ea \ee
Unlike the case of two string modes, the second term does not generally vanish. So because of the $\sqrt{2}$ factor, we can not combine the sums into a nice formula like (\ref{normalization}).

\section{Conclusion}  \label{secdiscussion} 

The $SO(8)$ rotational symmetry of the transverse directions in the pp-wave background (\ref{ppwave}) is broken by the Ramond-Ramond flux into $SO(4)\times SO(4)$, where the bosonic string modes are described differently by covariant derivatives and scalar field insertions in the dual CFT. As such, it is reasonable to expect our proposed entries of pp-wave holographic dictionary, e.g. (\ref{newentry}, \ref{dictionary2}, \ref{correspondence}), to face some challenges with more than four distinct string modes as the infinite Ramond-Ramond flux in our setting shall separate the two types of string modes. However, it is rather surprising that even for the case of four string modes, the torus two-point function can be negative, so the probability interpretation may no longer valid. Of course, since the two-point function is always real and symmetric, the arguments in \cite{Huang:2019lso} are still valid that it can not be naively identified with a quantum transition amplitude on the string theory side, which would then violate fundamental principle of unitarity.  It would be interesting to provide a reasonable explanation, or improve the proposed holographic dictionary (\ref{newentry}) to include this case of four string modes. 

On the other hand, we confirm that the factorization formulas e.g. (\ref{dictionary2}) are still valid for the case of four string modes, while the comparison with cubic string vertex (\ref{correspondence}) is seen to be straightforwardly applied to any hypothetical number of string modes, not even restricted by the eight dimensions of transverse directions in the pp-wave background. 

We also discuss the situation with multiple string modes in the same direction. In this case the BMN operators are no longer ``near-BPS", and there are potentially large quantum corrections on the field theory side if one turns on the gauge coupling. We check that the mathematical structures in the factorization formula and comparison with cubic string vertex, e.g. (\ref{dictionary2}, \ref{correspondence}), are robust and remain valid in this situation as we stay in free gauge theory. However, the proposed probability interpretation (\ref{newentry}) again seems rather fragile and  further breaks down in the case of three string modes because of a problem with normalization, though it still holds up in the case of two string modes due to the decoupling of the zero mode with non-zero modes.

It is interesting to further explore aspects of the pp-wave holographic dictionary. For example, in the case of three string modes, the non-negativity of torus two-point functions can be shown by explicit calculations, where there are numerous degenerate cases to deal with separately. One may ask whether there is a universal formalism which can deal with all cases regardless of mode number degeneracy and may also generalize to higher genus $h\geq 2$.  It is also interesting to check whether the factorization formulas are still valid in the case of more than four string modes or further in a hypothetical situation of any number of (different) string modes. Without a significant improvement of mathematical tools, the calculations are much more complicated. In any case, it seems worthwhile to push forward with the laborious endeavor for the purpose of a better understanding of pp-wave holography. 

As mentioned in \cite{Huang:2019uue}, the probability interpretation of two-point function implies the string perturbation series is convergent. In this sense, the holographic higher genus calculations are not asymptotic perturbative expansions as familiar in most  examples of quantum theories, but may in principle provide exact complete string amplitudes valid for any string coupling.  If no new non-perturbative effect is discovered in the future, then perhaps we have luckily found a rare example of perturbatively complete string theory, at least for the case of two string modes and very likely also for the case of three distinct string modes pending more tests of non-negativity at higher genus $h\geq 2$. In the cases of four or more string modes, the torus two-point functions are no longer always non-negative.  One can nevertheless similarly follow the method in \cite{Huang:2019uue} to give an upper bound on the higher genus two-point functions and show that the genus expansions remain convergent. For small string coupling and two different sets of mode numbers, the torus contribution is dominant, so the total two-point function could certainly be negative and is no longer a probability distribution although they can be still similarly normalized to sum to unity.  It would be desirable to better understand the physical meaning of the two-point function on the string theory side of the correspondence in this situation.

\vspace{0.2in} {\leftline {\bf Acknowledgments}}

We thank Jun-Hao Li, Jian-xin Lu, Gao-fu Ren, Pei-xuan Zeng for helpful discussions. This work was supported in parts by the national Natural Science Foundation of China (Grants No.11675167, No.11947301 and No.12047502).

\appendix

\section{Some calculational details of the one-loop string integrals} \label{details} 

We will convert the formulas for one-loop string diagrams (\ref{Sformulas}) into the 10 types of integrals in (\ref{4moderesult}). In the calculations, some cases are simply related to others by a transformation $(m_i,n_i)\rightarrow (-n_i, -m_i)$. It is helpful to first list the action of the transformation on the integrals 
\be \ba \label{transform1}
& I_i {~~\rm invariant} , ~~~ i=1,4,5, 6, 9,10, \\
 & I_2\leftrightarrow I_3, ~~~~  I_7\leftrightarrow I_8. 
\ea \ee
We discuss the dissection of the multi-dimensional integral domain in many cases, and introduce some positive variables  $z$'s and $z^{\prime}$'s such that they sum to one.

\subsection{$S_1$ contribution} 
We assume integral variables $y_4^{\prime} >y_4$. The other case is related by switching $y_4^{\prime} \leftrightarrow y_4$ and the transformation $(m_i,n_i)\rightarrow (-n_i, -m_i)$.  We have $0<y_i +y_4^{\prime} -y_4 <2x$. In this case first we write $1-x=\int_0^{1-x}  d z_7 dz_8  \delta(z_7+z_8-(1-x))$. Then the variables $z_7, z_8$ do not appear in the exponent. There is always an argument $0$ with at least multiplicity two in the standard integral. We define $z_4=x-y_4^{\prime}$, $z_5=y_4$ and discuss various cases. 

\begin{enumerate} 

\item $x<y_i +y_4^{\prime} -y_4 <2x, i=1,2,3$.  The delta function constrains fix $y_i^{\prime} = y_i +y_4^{\prime}-y_4-x, i=1,2,3$. Without loss of generality we assume $y_1>y_2>y_3$.  We change integration variables $z_i=x-y_i, i=1,2,3$,  $x=z_3+z_4+z_5+ z_6$, $z_3=z_3^{\prime}+z_2$, $z_2=z_2^{\prime} +z_1$.  The integral is then 
\be \ba
& \int_0^1 dz_1 dz_2^{\prime} d z_3^{\prime} [\prod_{i=4}^8 dz_i ] \delta (z_1+z_2^{\prime} +z_3^{\prime} +\sum_{i=4}^8 z_i -1)  \\
& \times e^{2\pi i [m_4 (z_4  + z_6) + n_4(z_1+z_5) +(n_4+n_1 -m_1) z_2^{\prime} + (m_3+m_4 -n_3)z_3^{\prime}  ]}    \\
& = I_{(2,2,2,1,1)} (m_4, n_4, 0, m_3+m_4 -n_3, n_4+n_1-m_1) .
\ea \ee
This is a $I_3$ type integral.

\item $x<y_i +y_4^{\prime} -y_4 <2x, i=2,3$, and $0<y_1 +y_4^{\prime} -y_4 <x$. The delta functions constrain $y_i^{\prime} = y_i +y_4^{\prime}-y_4-x, i=2,3$, and $ y_1^{\prime} = y_1+y_4^{\prime}-y_4$. 
Without loss of generality we assume $y_2>y_3$. We change variables $z_i=x-y_i, i=2,3, x=z_3+z_4+z_5+z_6, z_3=z_3^{\prime}+z_2$. We have $y_1<z_4+z_5$ and this further divides into two sub-cases.

\begin{enumerate}
\item $y_1<z_5$. Then we define $y_1=z_1, z_5=z_5^{\prime} +z_1$. The delta function constrain is $\delta(z_1+z_2 + z_3^{\prime}+ z_4 +z_5^{\prime} +z_6 -x)$. The exponents is now 
\be
e^{2\pi i [(n_1+n_4) (z_1+ z_2) +(m_1+m_4) (z_4+z_6 ) +(-n_3-m_2) z_3^{\prime} +(m_1+n_4) z_5^{\prime} ]}.
\ee
The integral is $I_{(2,2,2,1,1)}( m_1+m_4, n_1+n_4, 0, m_1+n_4, -n_3-m_2)$, which is a $I_4$ type integral.
   
\item $z_5<y_1<z_4+z_5$. Then we define $z_1=y_1-z_5, z_4=z_4^{\prime}+z_1$. The delta function constrain is $\delta(z_1+z_2 + z_3^{\prime}+ z_4^{\prime} +z_5 +z_6 -x)$. The exponents is  
\be
e^{2\pi i[(n_1+n_4) (z_2+ z_5) +(m_1+m_4) (z_4^{\prime}+z_6) +(-n_3-m_2) z_3^{\prime} +(m_4+n_1) z_1]}.
\ee
The integral is $I_{(2,2,2,1,1)}( m_1+m_4, n_1+n_4, 0, m_4+n_1, -n_3-m_2)$, which is also a $I_4$ type integral.   
   
\end{enumerate} 

\item $x<y_3+y_4^{\prime} -y_4 <2x$, and $0<y_i +y_4^{\prime} -y_4 <x, i=1,2$. The delta functions constrain $y_3^{\prime} = y_3+y_4^{\prime}-y_4-x$, and $ y_i^{\prime} = y_i+y_4^{\prime}-y_4, i=1,2$.  Without loss of generality we assume $y_1<y_2$. Define variables $z_3=x-y_3, x=z_3+z_4+z_5+z_6$. We have $y_i<z_4+z_5, i=1,2$ and this further divides into three sub-cases. 

\begin{enumerate}
\item $y_1<y_2<z_5$. Then we define $y_1=z_1, z_2=y_2-y_1, z_5=z_5^{\prime} +z_1+z_2$. The delta function constrain is 
$\delta(z_1+z_2 + z_3+ z_4 +z_5^{\prime} +z_6 -x)$. The exponents is 
\be
e^{2\pi i [-n_3(z_1+ z_3) -m_3 (z_4+z_6 ) + (-n_3+m_1-n_1)z_2 + (-m_3+n_4-m_4) z_5^{\prime}    ]}.
\ee
The integral is $I_{(2,2,2,1,1)}(-m_3, -n_3, 0, -m_3+n_4-m_4, -n_3 +m_1-n_1)$, which is a $I_2$ type integral. 

\item $y_1<z_5<y_2<z_4+z_5$. Then we define $y_1=z_1, z_2=y_2-z_5, z_5=z_5^{\prime} +z_1, z_4=z_2+z_4^\prime$. The delta function constrain is 
$\delta(z_1+z_2 + z_3+ z_4^\prime +z_5^{\prime} +z_6 -x)$. The exponents is 
\be
e^{2\pi i [-n_3(z_1+ z_3) -m_3 (z_4^\prime+z_6 ) + (-m_3+n_2-m_2)z_2 + (-n_3+m_1-n_1) z_5^{\prime}    ]}.
\ee
The integral is $I_{(2,2,2,1,1)}(-m_3, -n_3, 0, -m_3+n_2-m_2 , -n_3 +m_1-n_1)$, which is also a $I_2$ type integral.

\item $z_5<y_1<y_2<z_4+z_5$. Then we define $z_1=y_1-z_5, z_2=y_2-y_1,  z_4=z_1+z_2+z_4^\prime$. The delta function constrain is 
$\delta(z_1+z_2 + z_3+ z_4^\prime +z_5 +z_6 -x)$. The exponents is 
\be
e^{2\pi i [-n_3(z_3+ z_5) -m_3 (z_4^\prime+z_6 ) + (-m_3+n_2-m_2)z_2 + (-n_3+m_4-n_4) z_1   ]}.
\ee
The integral is $I_{(2,2,2,1,1)}(-m_3, -n_3, 0, -m_3+n_2-m_2 , -n_3 +m_4-n_4)$, which is also a $I_2$ type integral.

\end{enumerate} 

\item  $0<y_i +y_4^{\prime} -y_4 <x, i=1,2, 3$. The delta functions constrain $ y_i^{\prime} = y_i+y_4^{\prime}-y_4, i=1,2, 3$.  Without loss of generality we assume $y_1<y_2<y_3$. Define variables $x=z_4+z_5+z_6$. We have $y_i<z_4+z_5, i=1,2,3$ and this further divides into four sub-cases. 

\begin{enumerate}
\item $y_1<y_2<y_3<z_5$. Then we define $z_1=y_1, z_2=y_2-y_1, z_3=y_3-y_2, z_5=z_5^{\prime} +z_1+z_2+z_3$. The delta function constrain is 
$\delta(z_1+z_2 + z_3+ z_4 +z_5^{\prime} +z_6 -x)$. The exponents is 
\be
e^{2\pi i [ (m_1-n_1) z_2 + (n_3+n_4 -m_3-m_4) z_3 + (-m_4+n_4) z_5^\prime ]}.
\ee
The integral is $I_{(5,1, 1,1)}(0 , m_1-n_1, -m_4+n_4, n_3+n_4 -m_3-m_4 )$, which is a $I_1$ type integral. 

\item $y_1<y_2<z_5<y_3< z_4+z_5 $. Then we define $z_1=y_1, z_2=y_2-y_1, z_3=y_3-z_5, z_5=z_5^{\prime} +z_1+z_2, z_4=z_4^\prime +z_3 $. The delta function constrain is 
$\delta(z_1+z_2 + z_3+ z_4^\prime +z_5^{\prime} +z_6 -x)$. The exponents is 
\be
e^{2\pi i [ (m_1-n_1) z_2 + (-m_3+n_3) z_3 + (n_3+n_4 -m_3-m_4) z_5^\prime  ]}.
\ee
The integral is $I_{(5,1, 1,1)}(0 , m_1-n_1, -m_3+n_3, n_3+n_4 -m_3-m_4 )$, which is also a $I_1$ type integral. 

\item $y_1<z_5<y_2<y_3< z_4+z_5 $. Then we define $z_1=y_1, z_2=y_2-z_5, z_3=y_3-y_2, z_5=z_5^{\prime} +z_1, z_4=z_4^\prime +z_2+z_3 $. The delta function constrain is 
$\delta(z_1+z_2 + z_3+ z_4^\prime +z_5^{\prime} +z_6 -x)$. The exponents is 
\be
e^{2\pi i [ (m_1-n_1) z_5^\prime + (-m_3+n_3) z_3 + (n_2+n_3 -m_2-m_3) z_2  ]}.
\ee
The integral is $I_{(5,1, 1,1)}(0 , m_1-n_1, -m_3+n_3, n_2+n_3 -m_2-m_3 )$, which is also a $I_1$ type integral. 

\item $z_5<y_1< y_2<y_3< z_4+z_5 $. Then we define $z_1=y_1-z_5, z_2=y_2-y_1, z_3=y_3-y_2, z_4=z_4^\prime +z_1+z_2+z_3 $. The delta function constrain is 
$\delta(z_1+z_2 + z_3+ z_4^\prime +z_5 +z_6 -x)$. The exponents is 
\be
e^{2\pi i [ (m_4-n_4) z_1 + (-m_3+n_3) z_3 + (n_2+n_3 -m_2-m_3) z_2  ]}.
\ee
The integral is $I_{(5,1, 1,1)}(0 , m_4-n_4, -m_3+n_3, n_2+n_3 -m_2-m_3 )$, which is also a $I_1$ type integral. 

\end{enumerate} 

Summarizing the total contributions, taking into account various permutations of indices, we find 
\be 
S_1 = g^2(2I_1 + I_2 + I_3 +I_4). 
\ee

\end{enumerate}

\subsection{$S_2$ contribution} 

We only need to consider the first expression for $S_2$ in  (\ref{Sformulas}), and the others can be simply obtained by permutations of indices. First we consider the integrals of $y_4, y_4^{\prime}$. There are two cases 
\begin{enumerate} 
\item $y_4^\prime >y_4$. We define variables $z_4=y_4-x, z_4^\prime=y_4^\prime -y_4, z_5 =1-y_4^\prime $. There is a delta function constrain $\delta(z_4+z_4^\prime +z_5+x-1)$. The exponents of  $y_4, y_4^{\prime}$ variables become
\be  \label{S2case1} 
e^{2 \pi i [ (0) z_5 + n_4 z_4\prime +(n_4-m_4) z_4 +(n_4-m_4) x]  }  . 
\ee

\item $y_4^\prime <y_4$. This is simply obtained from the above by switching $n_4\rightarrow -m_4, m_4\rightarrow -n_4$. delta function constrain $\delta(z_4+z_4^\prime +z_5+x-1)$ is the same. The exponent is now 
\be  \label{S2case2} 
e^{2 \pi i [ (0) z_5 - m_4 z_4\prime +(n_4-m_4) z_4 +(n_4-m_4) x]  }  . 
\ee

\end{enumerate}

Next we consider the integrals of $y_i, y_i^\prime, i=1,2,3$. We assume $y_3^\prime >y_3$, with the other cases obtained by the transformation (\ref{transform1}). We have $0<y_i +y_3^\prime -y_3 <2x, i=1,2,$. We define $z_3^\prime =x-y_3, z_3=y_3$ and discuss various cases 

\begin{enumerate}

\item $x<y_i + y_3^\prime -y_3 <2x, i=1,2$. The delta functions constrain $ y_i^{\prime} = y_i+y_3^{\prime}-y_3-x, i=1,2$.  Without loss of generality we assume $y_1>y_2$. Define variables $z_2=z_2^\prime+z_1,  x=z_1+z_2^\prime+z_3+ z_3^\prime+z_6$. Including the factor $e^{2 \pi i (n_4-m_4) x} $, the exponents of $y_i, y_i^{\prime}, i=1,2,3$ variables become
\be 
e^{2\pi i [(n_3+n_4-m_4) z_1 + (-n_2-m_1-m_4) z_2^\prime + (n_3+n_4) z_3 + (m_3+m_4) z_3^\prime +m_3 z_6 ] }. 
\ee
There are two contributions. Combining with equation (\ref{S2case1}) we have an integral 
\be
I(m_3, n_3, -m_4, -n_4, -m_4-n_4, m_3-n_4, n_3-m_4, m_3+n_3+m_2+n_1), 
\ee
which is a $I_7$ type integral, while combining with equation (\ref{S2case2}) we have an integral 
\be
I(m_3, n_3, -m_4, -n_4, 0, m_3-n_4, n_3-m_4, n_3-m_3 +n_1-m_1), 
\ee
which is a $I_6$ type integral.

\item $x<y_2+ y_3^\prime -y_3 <2x, 0<y_1+ y_3^\prime -y_3 <x$. The delta functions constrain $ y_2^{\prime} = y_2+y_3^{\prime}-y_3-x, y_1^{\prime} = y_1+y_3^{\prime}-y_3 $.  Define variables $z_2=x-y_2,  x=z_2+z_3+ z_3^\prime+z_6$.  We have $y_1<z_3+z_3^\prime$, and discuss two sub-cases 

\begin{enumerate} 
\item $y_1<z_3$. Define variables $z_1=y_1, z_3=z_1+z_1^\prime$. Including the factor $e^{2 \pi i (n_4-m_4) x} $,  the exponents of $y_i, y_i^{\prime}, i=1,2,3$ variables become
\be 
e^{2\pi i [ -n_2 z_1 + (m_1 + n_3 +n_4 ) z_1^\prime + (-m_4 - n_2 ) z_2 + (-m_2-m_4+n_4) z_3^\prime +(-m_2-m_4) z_6 ] }. 
\ee
There are two contributions. Combining with equation (\ref{S2case1}) we have an integral 
\be
I(m_4, n_4, -m_2, -n_2, m_4+n_4, m_4-n_2, n_4-m_2, m_4+n_4+m_1+n_3), 
\ee
which is a $I_8$ type integral, while combining with equation (\ref{S2case2}) we have an integral 
\be
I(m_4, n_4, -m_2, -n_2, 0,  m_4-n_2, n_4-m_2, m_4+n_4+m_1+n_3), 
\ee
which is a $I_6$ type integral.

\item $z_3<y_1<z_3+z_3^\prime$. Define variables $z_1=y_1-z_3, z_3^\prime=z_1+z_1^\prime$. Including the factor $e^{2 \pi i (n_4-m_4) x} $,  the exponents of $y_i, y_i^{\prime}, i=1,2,3$ variables become
\be 
e^{2\pi i [ (m_3+n_1+n_4) z_1 + (-m_2-m_4+n_4) z_1^\prime + (-m_4 -n_2) z_2 -n_2 z_3 +(-m_2-m_4) z_6 ] }. 
\ee
There are two contributions. Combining with equation (\ref{S2case1}) we have an integral 
\be
I(m_4, n_4, -m_2, -n_2, m_4+n_4, m_4-n_2, n_4-m_2, m_4+n_4+m_3+n_1), 
\ee
which is a $I_8$ type integral, while combining with equation (\ref{S2case2}) we have an integral 
\be
I(m_4, n_4, -m_2, -n_2, 0,  m_4-n_2, n_4-m_2, m_4+n_4+m_3+n_1), 
\ee
which is a $I_6$ type integral. 

\end{enumerate} 

\item $0<y_i+ y_3^\prime -y_3 <x, i=1,2 $. The delta functions constrain $ y_i^{\prime} = y_i+y_3^{\prime}-y_3 , i=1,2$.  Without loss of generality we assume $y_1<y_2$. Define variables $x=z_3+ z_3^\prime+z_6$.  We have $y_i<z_3+z_3^\prime, i=1,2$, and discuss three sub-cases 

\begin{enumerate}

\item $y_1<y_2<z_3$. Define variables $z_1=y_1, z_2 = y_2-y_1, z_3=y_2 +z_2^\prime$. Including the factor $e^{2 \pi i (n_4-m_4) x} $,  the exponents of $y_i, y_i^{\prime}, i=1,2,3$ variables become
\be 
e^{2\pi i [ (0) z_1 + (m_1 - n_1 ) z_2 + (n_3+n_4 -m_3-m_4  ) z_2^\prime + (n_4 -m_4 ) z_3^\prime  -m_4 z_6 ] }. 
\ee
There are two contributions. Combining with equation (\ref{S2case1}) we have an integral 
\be
I_{(2,2,1,1,1,1)} (n_4 -m_4, 0, n_4, -m_4, m_1-n_1, n_4-m_4+n_3-m_3) ,  
\ee
which is a $I_5$ type integral, while combining with equation (\ref{S2case2}) we have an integral 
\be
I_{(2,2,2,1,1)} (m_4, n_4, 0, n_4+n_3-m_3, m_4+m_1-n_1 ), 
\ee
which is a $I_3$ type integral. 

\item $y_1<z_3<y_2<z_3+z_3^\prime$. Define variables $z_1=y_1, z_2 = y_2-z_3, z_3=z_1 +z_1^\prime, z_3^\prime =z_2+z_2^\prime$. Including the factor $e^{2 \pi i (n_4-m_4) x} $,  the exponents of $y_i, y_i^{\prime}, i=1,2,3$ variables become
\be 
e^{2\pi i [ (0) z_1 + (m_1 - n_1 ) z_1^\prime + (n_2+n_4 -m_2-m_4  ) z_2 + (n_4 -m_4 ) z_2^\prime  -m_4 z_6 ] }. 
\ee
There are two contributions. Combining with equation (\ref{S2case1}) we have an integral 
\be
I_{(2,2,1,1,1,1)} (n_4 -m_4, 0, n_4, -m_4, m_1-n_1, n_4-m_4+n_2-m_2) ,  
\ee
which is a $I_5$ type integral, while combining with equation (\ref{S2case2}) we have an integral 
\be
I_{(2,2,2,1,1)} (m_4, n_4, 0, n_4+n_2-m_2, m_4+m_1-n_1 ), 
\ee
which is a $I_3$ type integral.

\item $z_3<y_1<y_2<z_3+z_3^\prime$. Define variables $z_1=y_1-z_3, z_2 = y_2-y_1, z_3^\prime =z_1+z_2+z_2^\prime$. Including the factor $e^{2 \pi i (n_4-m_4) x} $,  the exponents of $y_i, y_i^{\prime}, i=1,2,3$ variables become
\be 
e^{2\pi i [ (m_3-n_3) z_1 + (n_2+n_4 -m_2-m_4  ) z_2 + (n_4 -m_4 ) z_2^\prime +(0)z_3 -m_4 z_6 ] }. 
\ee
There are two contributions. Combining with equation (\ref{S2case1}) we have an integral 
\be
I_{(2,2,1,1,1,1)} (n_4 -m_4, 0, n_4, -m_4, m_3-n_3, n_4-m_4+n_2-m_2) ,  
\ee
which is a $I_5$ type integral, while combining with equation (\ref{S2case2}) we have an integral 
\be
I_{(2,2,2,1,1)} (m_4, n_4, 0, n_4+n_2-m_2, m_4+m_3-n_3 ), 
\ee
which is a $I_3$ type integral.

\end{enumerate} 
 
\end{enumerate} 

Summarizing the total contributions, taking into account various permutations of indices, we find 
\be 
S_2 = g^2( I_2 + I_3 +2 I_5 +2I_6+I_7+I_8). 
\ee

\subsection{$S_3$ contribution} 

We only need to consider the first expression for $S_3$ in  (\ref{Sformulas}), and the others can be simply obtained by permutations of indices. First we consider the integrals of $y_i, y_i^{\prime}, i=3,4$. We assume $y_4^\prime >y_4$, with the other cases obtained by the transformation (\ref{transform1}).  Define variables $z_4=y_4-x, z_4^\prime =1 - y_4^\prime $. We have $x<y_3+y_4^\prime -y_4<2-x$. There are two cases with a subdivision into a total of three cases 

\begin{enumerate} 
\item $1<y_3+y_4^\prime -y_4<2-x$. The delta function constrain $y_3^\prime = y_3 +y_4^\prime -y_4-(1-x)$. We define variable $y_3=1-z_3, 1-x = z_3+z_4+z_4^\prime +z_5$. The exponents of  $y_i, y_i^{\prime}, i=3,4$ variables become
\be  \label{S3case1} 
e^{2 \pi i [ n_4z_3 +(n_4-m_3-m_4)z_4 -m_3z_4^\prime + (n_3+n_4-m_3)z_5  +(n_3+n_4-m_3-m_4) x]  }  . 
\ee

\item $x<y_3+y_4^\prime -y_4<1$. The delta function constrain $y_3^\prime = y_3 +y_4^\prime -y_4$. We have $y_3-x<z_4+z_4^\prime$, which divides into two sub-cases 
\begin{enumerate}
\item $y_3-x<z_4$. Define variables $z_3=y_3-x, z_4=z_3+z_3^\prime$. The exponents of  $y_i, y_i^{\prime}, i=3,4$ variables become
\be  \label{S3case2} 
e^{2 \pi i [ (n_3+n_4 -m_3 -m_4) z_3 +(n_4-m_4)z_3^\prime  +(0) z_4^\prime + (n_3+n_4)z_5  +(n_3+n_4-m_3-m_4) x]  }  . 
\ee

\item $z_4<y_3-x<z_4+z_4^\prime $. Define variables $z_3=y_3-x-z_4 , z_4^\prime =z_3+z_3^\prime$. The exponents of  $y_i, y_i^{\prime}, i=3,4$ variables become
\be  \label{S3case3} 
e^{2 \pi i [ (n_3 -m_3 ) z_3 +(0 )z_3^\prime  +(n_3+n_4-m_3-m_4 ) z_4^\prime + (n_3+n_4)z_5  +(n_3+n_4-m_3-m_4) x]  }  . 
\ee

\end{enumerate} 

\end{enumerate} 

Next we consider the integrals of $y_i, y_i^{\prime}, i=1,2$. We mainly consider $y_2^\prime>y_2$ and the results for $y_2^\prime<y_2$ can be simply obtained by transforming $(m_i, n_i)\rightarrow (-n_i, -m_i), i=1,2$. We define $z_2^\prime = x-y_2^\prime, z_2=y_2$. We have $0<y_1+y_2^\prime -y_2<2x$ and discuss some cases 

\begin{enumerate} 

\item $x<y_1+y_2^\prime -y_2<2x$. The delta function constrain $y_1^\prime = y_1 +y_2^\prime -y_2-x $.  Define $y_1=x-z_1, x=z_1+z_2+z_2^\prime +z_6$.  The exponents of $y_i, y_i^{\prime}, i=1,2$ variables become
\be \label{A37}
e^{2\pi i [ n_2 z_1 + (n_2 -m_1 -m_2) z_2 -m_1 z_2^\prime + (n_1+n_2 -m_1) z_6 ] }. 
\ee
There are three contributions. Combining with equation (\ref{S3case1}) we have an integral 
\be
I (m_2, n_2, -m_3, -n_3, m_1+m_2+n_2, n_1+n_2+m_2, m_1+m_2-n_3, n_1+n_2-m_3) ,  
\ee
which is a $I_{10}$ type integral, combining with equation (\ref{S3case2}) we have an integral 
\be
I ( m_1, n_1, -m_2, -n_2, -m_2-n_2, m_1-n_2, n_1-m_2, m_1+n_1+m_3+n_4 ), 
\ee
which is a $I_7$ type integral, and combining with equation (\ref{S3case3}) we have an integral 
\be
I ( m_1, n_1, -m_2, -n_2, -m_2-n_2, m_1-n_2, n_1-m_2, m_1+n_1+m_4+n_3 ), 
\ee
which is also a $I_7$ type integral.

\item We transform equation (\ref{A37}) by $(m_i, n_i)\rightarrow (-n_i, -m_i), i=1,2$, and get a factor  
\be
e^{2\pi i [ -m_2 z_1 + (-m_2 +n_1 +n_2) z_2 +n_1 z_2^\prime + (-m_1-m_2 +n_1) z_6 ] }. 
\ee
Again there are three contributions. Combining with equation (\ref{S3case1}) we have an integral 
\be
I (m_4, n_4, -m_2, -n_2, m_3+m_4+n_4, n_3+n_4+m_4, m_3+m_4-n_2, n_3+n_4-m_2) ,  
\ee
which is a $I_{10}$ type integral, combining with equation (\ref{S3case2}) we have an integral 
\be
I ( m_2, n_2, -m_1, -n_1, -m_1-n_1, m_2-n_1, n_2-m_1, m_2+n_2+m_3+n_4 ), 
\ee
which is a $I_7$ type integral, and combining with equation (\ref{S3case3}) we have an integral 
\be
I ( m_2, n_2, -m_1, -n_1, -m_1-n_1, m_2-n_1, n_2-m_1, m_2+n_2+m_4+n_3 ) , 
\ee
which is also a $I_7$ type integral.

\item $0<y_1+y_2^\prime -y_2<x$. The delta function constrain $y_1^\prime = y_1 +y_2^\prime -y_2 $.  We have $y_1<z_2^\prime +z_2$ and discuss some sub-cases  

\begin{enumerate} 

\item $0<y_1<z_2$. Define $z_1=y_1, z_2=z_1+z_1^\prime, x=z_1+z_1^\prime +z_2^\prime +z_6$.  The exponents of $y_i, y_i^{\prime}, i=1,2$ variables become
\be \label{A45}
e^{2\pi i [ (n_1+n_2-m_1-m_2)z_1 + (n_2-m_2) z_1^\prime +(0) z_2^\prime +(n_1+n_2)z_6  ] }. 
\ee
There are three contributions. Combining with equation (\ref{S3case1}) we have an integral 
\be
I (m_3, n_3, -m_4, -n_4, -m_4-n_4, m_3-n_4, n_3-m_4, m_3+n_3 +m_1+n_2) ,  
\ee
which is a $I_{7}$ type integral, combining with equation (\ref{S3case2}) we have an integral 
\be
I _{(2,2,1,1,1,1)} (n_3+n_4 -m_3-m_4, 0, n_3+n_4, -m_3-m_4, n_4-m_4, -n_1+m_1  ), 
\ee
which is a $I_9$ type integral, and combining with equation (\ref{S3case3}) we have an integral 
\be
I _{(2,2,1,1,1,1)} (n_3+n_4 -m_3-m_4, 0, n_3+n_4, -m_3-m_4, n_3-m_3, -n_1+m_1  ),   
\ee
which is also a $I_9$ type integral.

\item  We transform equation (\ref{A45}) by $(m_i, n_i)\rightarrow (-n_i, -m_i), i=1,2$, and get a factor  
\be
e^{2\pi i [ (n_1+n_2-m_1-m_2)z_1 + (n_2-m_2) z_1^\prime +(0) z_2^\prime +(-m_1-m_2)z_6  ] }.  
\ee
Again there are three contributions. Combining with equation (\ref{S3case1}) we have an integral 
\be
I (m_3, n_3, -m_4, -n_4, m_3+n_3, m_3-n_4, n_3-m_4, m_3+n_3 + m_1+n_2 ) ,  
\ee
which is a $I_{8}$ type integral, combining with equation (\ref{S3case2}) we have an integral 
\be
I _{(2,2,2,1,1)} (m_1+m_2, n_1+n_2, 0, m_1+n_2, -m_4-n_3 ), 
\ee
which is a $I_4$ type integral, and combining with equation (\ref{S3case3}) we have an integral 
\be
I _{(2,2,2,1,1)} ( m_1+m_2, n_1+n_2, 0, m_1+n_2, -m_3-n_4 ),   
\ee
which is also a $I_4$ type integral.

\item $z_2<y_1<z_2+z_2^\prime$. Define variables $z_1=y_1-z_2, z_2^\prime =z_1+z_1^\prime, x= z_1+z_1^\prime +z_2 +z_6$. The exponents of $y_i, y_i^{\prime}, i=1,2$ variables become
\be 
e^{2\pi i [ (n_1+n_2-m_1-m_2)z_2 + (n_1-m_1) z_1 +(0) z_1^\prime +(n_1+n_2)z_6  ] }. 
\ee
We notice this is just (\ref{A45}) with the index switch $1\leftrightarrow 2$. So we can simply obtain the remaining results by index switching the last two sub-cases. Namely we have one more type of $I_7, I_8$ integrals, two more types of $I_4, I_9$ integrals.

\end{enumerate} 

\end{enumerate} 

Summarizing the total contributions, taking into account various permutations of indices, we find 
\be 
 S_3 = g^2(I_4 + +I_7 +I_8 +2I_9 +2I_{10} ). 
\ee

\addcontentsline{toc}{section}{References}

%\bibliographystyle{utphys} 
%\bibliography{Reference1}

\providecommand{\href}[2]{#2}\begingroup\raggedright\endgroup

\end{document}